\documentclass[preprint,prf,a4paper,superscriptaddress,longbibliography]{revtex4-1}
\usepackage{graphicx,color,epsfig,subfig,dcolumn,bm,mathrsfs,amsmath}
\usepackage[T1]{fontenc}

\newcommand{\ud}{\mathrm{d}}
\newcommand{\uD}{\mathrm{D}}

\DeclareMathAlphabet{\mathsfit}{\encodingdefault}{\sfdefault}{m}{sl}
\SetMathAlphabet{\mathsfit}{bold}{\encodingdefault}{\sfdefault}{bx}{sl}
\newcommand{\tens}[1]{\bm{\mathsfit{#1}}}

\begin{document}

\title{Dynamics of Viscoelastic Filaments Based on Onsager Principle}    

\author{Jiajia Zhou}
\email[]{jjzhou@buaa.edu.cn}
\affiliation{Key Laboratory of Bio-Inspired Smart Interfacial Science and Technology of Ministry of Education, School of Chemistry, Beihang University, Beijing 100191, China} 
\affiliation{Center of Soft Matter Physics and Its Applications, Beihang University, Beijing 100191, China}
\author{Masao Doi}
\email[]{masao.doi@buaa.edu.cn}
\affiliation{Center of Soft Matter Physics and Its Applications, Beihang University, Beijing 100191, China}

% \date{\today}

\begin{abstract}
When a polymer solution is uniaxially stretched and held fixed at both ends, the solution quickly separates into droplets connected by strings and takes the beads-on-string structure. 
The string then becomes thinner by capillary forces. 
Here we develop a theoretical framework on viscoelastic fluids based on Onsager principle, and apply it to the dynamics of viscoelastic filaments. 
We show that the beads-on-string structure is a thermodynamic quasi-equilibrium state, and derive an equation for the coexistence condition in the pseudo-equilibrium state. 
Using the condition, we solve the evolution equation analytically and show that the string radius and the tensile stress vary exponentially as predicted by the classical theory of Entov and Hinch [J. Non-Newtonian Fluid Mech. {\bf 72}, 31 (1997)], but the prefactor for the tensile stress is different from their theory and agrees with the numerical solutions of Clasen et al. [J. Fluid Mech. {\bf 556}, 283 (2006)].  
%This justifies the validity of the coexistence condition obtained by the Onsager principle.
\end{abstract}

% \pacs{}

\maketitle

%%%%%%%%%%%%%%%%%%%%%%%%%%%%%%%%%%%%%%%%%%%%%%%%%%%%%%%%%%%%%%
\section{Introduction}

Viscoelastic fluids exhibit many interesting flow phenomena which do not appear in Newtonian fluids \cite{BAH1,BAH2}. 
These flow properties have found many industrial applications such as fiber spinning, film blowing, and jet printing \cite{Yarin_fibers}. 
One particular phenomenon that has been studied extensively is the necking and breakup of viscoelastic fluids under extensional flow \cite{Yarin, McKinley2002, McKinley2005, HuangQian2017}.
A filament of Newtonian fluid is unstable due to Plateau-Rayleigh instability, and the fluid column rapidly separates into droplets \cite{Eggers1997}. 
The situation becomes quite different when small amount of high molecular weight polymers are added. 
The viscoelastic filament collapses into droplets connected by uniform strings (\emph{beads-on-string} structure) \cite{Goldin1969, Renardy1994, Renardy1995, LiJie2003, Wagner2005, Clasen2006, Sattler2008}.
The strings remain stable and continue thinning exponentially by the surface tension \cite{Entov1997}.

Previous studies on viscoelastic fluids are mostly based on prescribed constitutive equations. 
In this paper, we take a different approach based on thermodynamics \cite{Beris_Edwards, Oettinger_beyond} and derive evolution equations using Onsager principle \cite{DoiSoft}.
We first show that the dynamics of viscoelastic fluids can be cast into the general framework of Onsager principle. 
We then apply Onsager principle to study the dynamics of viscoelastic filaments. 
We start from a free energy model based on the filament geometry and microscopic chain extension.
Together with a dissipation mechanism which accounts relative motion between the polymer and the fluid, we derive evolution equations of a viscoelastic filament.
This framework interprets formation of the beads-on-string structure from the perspective of phase separation.
We further analyze the elasto-capillary thinning and derive the exponential thinning for Oldroyd-B model.
Different from the classical work of Entov and Hinch \cite{Entov1997}, who made an assumption on the axial stress, our work is based on the dynamical coexistence between the string and the bead.
We will show that this gives the same results as the full numerical solutions by Clasen et al. \cite{Clasen2006}.

%%%%%%%%%%%%%%%%%%%%%%%%%%%%%%%%%%%%%%%%%%%%%%%%%%%%%%%
\section{Onsager variational principle}
\label{sec:Onsager}

%======================================
\subsection{Onsager principle}
Onsager principle is a variational principle proposed by Onsager in his celebrated paper on the reciprocal relation \cite{Onsager1931}.  
He showed that because of the reciprocal relation, the evolution law in irreversible systems can be stated in a form of variational principle.  
For isothermal system, the variational principle has the same form as the Rayleigh's minimum energy dissipation principle in hydrodynamics. 
Let  $x=(x_1, x_2, \cdots)$ be the set of variables that characterizes the nonequilibrium state of a system. 
Then the time evolution of the system is determined by the condition that the following quadratic function of $\dot{x}=(\dot{x}_1, \dot{x}_2, \cdots)$ be minimized with respect to $\dot{x}$,
\begin{equation}
 \mathscr{R}=\frac{1}{2}  \sum_{i,\,j} \zeta_{ij} \dot{x}_i \dot{x}_j 
          +   \sum_i \frac{\partial A}{\partial x_i} \dot{x}_i
                             \label{eqn:D1}
\end{equation}
where $A(x)$ is the free energy of the system, and $\zeta_{ij}$ is a function of $x$, corresponding to the friction coefficient in hydrodynamics.  
The minimum condition gives
\begin{equation}
  \sum_j \zeta_{ij} \dot{x}_j =- \frac{\partial A}{\partial x_i}
                             \label{eqn:D2}
\end{equation}
If $\dot{x}$ is taken to be the time derivative of $x$, Eq.~(\ref{eqn:D2}) represents the time evolution equation for $x$. 
Equation (\ref{eqn:D2}) is an analogue to the force balance equations in Stokesian hydrodynamics, and is called Onsager's kinetic equation.
The reciprocal relation $\zeta_{ij}=\zeta_{ji}$ is guaranteed in this derivation.

Many equations used in soft matter systems, such as Stokes equation, diffusion equation, Nernst-Planck equation, Cahn-Hilliard equation, Ericksen-Leslie equation, etc., have been derived based on Onsager principle \cite{DoiSoft, Doi2012}.
In the following, we shall show that the continuum mechanical equation for  viscoelastic fluid can also be derived from the Onsager principle.
There are various models (or constitutive equations) for viscoelastic fluid.  
Here we focus on the Oldroyd-B model. 
Other viscoelastic models can be treated in the similar manner.

%============================================================
\subsection{Constitutive equation of Oldroyd-B fluid}
The Oldroyd-B model is a model describing the mechanical property of viscoelastic fluid. 
It gives a mathematical relation between the stress tensor $\tens{\sigma}$ and the 
velocity gradient tensor $\tens{\kappa}$, defined by $ \kappa_{\alpha \beta} = \partial v_{\alpha}/\partial r_{\beta}$ ($\alpha, \beta=x,y,z$). 
 
The stress tensor of the Oldroyd-B model is written as
\begin{equation}
  \tens{\sigma} = G(\tens{c}-\tens{I}) 
                             + \eta_s (\tens{\kappa} + \tens{\kappa}^t) - p \tens{I} ,
                                         \label{eqn:D5}
\end{equation}
where $\tens{c}$ is a tensor describing the molecular deformation.
In the molecular theory, $\tens{c}$ is related to the average of the end-to-end vector $\mathbf{R}$ of the polymer chains by $\tens{c}=3 \langle \mathbf{R} \mathbf{R} \rangle / 
\langle \mathbf{R}^2 \rangle_{\rm eq}$, but here $\tens{c}$ can be understood as a dimensionless quantity obtained as a solution of the following time evolution equation
\begin{equation}
  \dot{ \tens{c} } - \tens{\kappa} \cdot \tens{c} 
          - \tens{c} \cdot \tens{\kappa}^t 
    = - \frac{1}{\tau} (\tens{c} -  \tens{I}).
  \label{eqn:D6}
\end{equation}
In Eqs.~(\ref{eqn:D5}) and (\ref{eqn:D6}), $G$, $\tau$ and $\eta_s$ are material parameters, representing the shear modulus, relaxation time and solvent viscosity.  
Equations~(\ref{eqn:D5}) and (\ref{eqn:D6}) together with the incompressible condition
\begin{equation}
  \nabla \cdot \mathbf{v} = 0 
  \label{eqn:D7}
\end{equation}
give the constitutive equation for the Oldroyd-B fluid.

The above set of equations can be formulated in the framework of Onsager principle. 
In Appendix \ref{app:details}, we derived the dissipation function for the Oldroyd-B model
\begin{equation}
     \Phi_p(\dot {\tens{c}}; \tens{c}) = \frac{G\tau}{4} {\rm Tr} 
          [ \tens{c}^{-1} ( \dot{\tens{c}}^t - \tens{\kappa} \cdot \tens{c} 
                               -  \tens{c} \cdot \tens{\kappa}^t) 
                                  ( \dot{\tens{c}} - \tens{\kappa} \cdot \tens{c} 
                                  - \tens{c} \cdot \tens{\kappa}^t) ] .                         
    \label{eqn:D8}
\end{equation}
The dissipation function accounts the friction due to the relative motion between the polymer segments and the fluid.

We also showed that the free energy for the $\tens{c}$-tensor has the form
\begin{equation}
  A_p (\tens{c}) = \frac{1}{2} G \, \big[ {\rm Tr} (\tens{c}) - \ln \det(\tens{c}) \big] .
  \label{eqn:D9}
\end{equation}
The time derivative of the free energy is calculated as
\begin{equation}
   \dot{A}_p(\dot {\tens{c}}; \tens{c}) = 
              \frac{1}{2} G \, {\rm Tr} \big[ ( \tens{I} - \tens{c}^{-1})
                 \cdot \dot{\tens{c}} \big] .
   \label{eqn:D10}
\end{equation}
The Rayleighian is then given by 
\begin{equation}
  \mathscr{R}_p(\dot{ \tens{c}}; \tens{c})  
              = \Phi_p (\dot{ \tens{c}}; \tens{c}) 
                                      + \dot{A}_p(\dot{ \tens{c}}; \tens{c}) .
  \label{eqn:D11}
\end{equation}
It is easy to check that the relation $\partial \mathscr{R}_p/\partial \dot{\tens{c}} = 0$ gives Eq.~(\ref{eqn:D6}).

One can also check that the polymer contribution to the stress tensor
$\tens{\sigma}_p = G (\tens{c}-\tens{I})$ can also be derived from the Rayleighian by
\begin{equation}
  (\sigma_{p})_{\alpha\beta}
     =\frac{\partial \mathscr{R}_p}{\partial \kappa_{\alpha\beta}} .
                               \label{eqn:D12}
\end{equation}
This relation will be used in the next subsection.

%==============================================
\subsection{Flow of Oldroyd-B fluid }
In the previous section, we considered the behavoir of Oldroyd-B fluid for a given velocity gradient $\tens{\kappa}$.  
We now consider how the Oldroyd-B fluid flows for given boundary conditions. 
The velocity field $\mathbf{v}(\mathbf{r})$ of the Oldroyd-B model can also be determined by the Onsager principle. 
For the sake of simplicity, here we consider the case that the inertial term (density times acceleration) can be ignored in the momentum balance equation.

If the inertia term is ignored, the velocity field $\mathbf{v}(\mathbf{r})$ is determined by the force balance equation
\begin{equation}
      \nabla \cdot \tens{\sigma} - \nabla p =0 .
                              \label{eqn:D13}  
\end{equation}
The constitutive equations (\ref{eqn:D5}), (\ref{eqn:D6}), (\ref{eqn:D7})  
and the force balance equation (\ref{eqn:D13}) determine the time evolution of $\tens{c}(\mathbf{r};t)$ and the velocity field $\mathbf{v}(\mathbf{r};t)$.  

One can show that these equations are derived from the Onsager principle by minimizing the following Rayleighian
\begin{equation}
       \mathscr{R}_{\rm tot}[\dot{ \tens{c}}, \mathbf{v}]  
       =  \int \ud \mathbf{r} \, [ \mathscr{R}_{p}(\dot{ \tens{c}},\tens{\kappa})
                                 + \mathscr{R}_{s}(\tens{\kappa})
                                 +p {\rm Tr} \tens{\kappa} ] .
                              \label{eqn:D14}  
\end{equation}
In Eq.~(\ref{eqn:D14}), $\mathscr{R}_{s}(\tens{\kappa})$ stands for the solvent contribution to the Rayleighian
\begin{equation}
      \mathscr{R}_{s}(\tens{\kappa})=\frac{1}{4} \eta_s 
      (\tens{\kappa} + \tens{\kappa}^t): (\tens{\kappa} + \tens{\kappa}^t) .
                                          \label{eqn:D15}  
\end{equation}
The last term in Eq.~(\ref{eqn:D14}) is introduced to account for the incompressible condition (\ref{eqn:D7}). 
Notice that the minimization must be done with respect to $\dot{\tens{c}}$ and $\mathbf{v}$. [Here $\dot{\tens{c}}$ should be understood as 
the material time derivative $\uD_t \tens{c}$, but this does 
not affect the variational calculus to minimize the Rayleighian 
(see Appendix \ref{app:1dEqn})].
The condition $\delta \mathscr{R}_{\rm tot} / \delta \dot{\tens{c}} =0$ gives the constitutive equation, and the condition $\delta \mathscr{R}_{\rm tot} / \delta \mathbf{v} = 0$ gives the force balance equation.

To summarize, the constitutive equation and the force balance equation for the Oldroyd-B model can be derived from the Onsager principle.
They are determined by minimizing the Rayleighian (\ref{eqn:D14}) with respect to $\dot{\tens{c}}$ and $\mathbf{v}$.

%%%%%%%%%%%%%%%%%%%%%%%%%%%%%%%%%%%%%%%%%%%%%%
\section{Uniaxially stretched filament}

%=================================
\subsection{Uniform stretching}
\label{sec:Uniform_stretching}

\begin{figure}[htbp]
  \centering
  \includegraphics[width=0.5\textwidth]{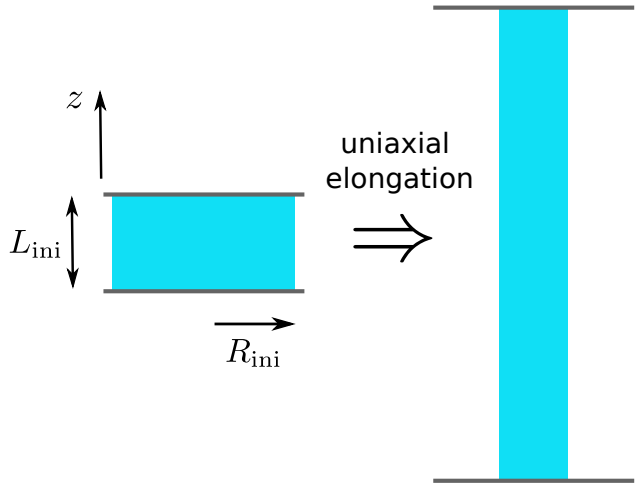}
  \caption{A cylindrical specimen of initial radius $R_{\rm ini}$ and length $L_{\rm ini}$ is uniformly stretched in the $z$-direction.}
  \label{fig:sketch1}
\end{figure}

As an example of the above approach, we consider a simple case of uniaxially stretched filament.
We consider that a filament with initial radius $R_{\rm ini}$ and length $L_{\rm ini}$ is stretched uniformly and has radius $R$ and length $L$(see Fig.~\ref{fig:sketch1}).  
The elongation ratio $\lambda$, and Hencky strain $\varepsilon$ are
defined by
\begin{eqnarray}
  \label{eq:LR}
  \lambda  &=& \frac{L}{L_{\rm ini}}= \frac{R_{\rm ini}^2}{R^2}  ,\\
  \varepsilon &=& \ln \lambda  ,
\end{eqnarray}
where the volume conservation condition 
 $ L_{\rm ini} R_{\rm ini}^2= LR^2$ has been used.

The velocity-gradient tensor is given by
\begin{equation}
 \tens{\kappa} = 
    \begin{pmatrix}
      \kappa_{xx} &  0 &  0 \\
       0 & \kappa_{yy} &  0 \\
       0 &  0 &  \kappa_{zz} 
     \end{pmatrix}
 =\frac{1}{2} \dot{\varepsilon}(t) 
    \begin{pmatrix}
      -1 &  0 &  0 \\
       0 & -1 &  0 \\
       0 &  0 &  2 
     \end{pmatrix}
    \label{eq:kappa0}
\end{equation}
where $\dot{\varepsilon}$ is the Hencky strain rate
($\dot{\varepsilon}= \dot \lambda/\lambda$).

In the present problem,  the $\tens{c}$-tensor is diagonal and 
can be written as
\begin{equation}
  \tens{c} = 
    \begin{pmatrix}
      c_{xx} &  0 &  0 \\
       0 & c_{yy} &  0 \\
       0 &  0 &  c_{zz} 
     \end{pmatrix} \, .
    \label{eq:ctensor0}
\end{equation} 
For system with cylindrical symmetry, we have $c_{xx}=c_{yy}=c_{rr}$ for the radial component, and $c_{zz}$ is the axial component.

The free energy of the filament per unit volume is then calculated as 
\begin{eqnarray}
  \frac{A}{V} &=&  \frac{1}{2} G(c_{xx} + c_{yy} + c_{zz} - \ln c_{xx}c_{yy}c_{zz} ) 
                       +\frac{2 \pi RL \gamma}{V} \\
              &=& \frac{1}{2} G \left[ (c_{xx} + c_{yy} + c_{zz} -\ln c_{xx}c_{yy}c_{zz} ) 
                          + 4 \alpha\sqrt{\lambda} \right],
  \label{eq:el_A1}
\end{eqnarray}
where the term including the surface tension $\gamma$ represents the surface energy contribution. 
Here a dimensionless number $\alpha$ (also called elasto-capillary number) appears, which is 
defined by $\alpha = \gamma/(G R_{\rm ini})$.

The dissipation function of the system is now given by [see Eqs.~(\ref{eqn:D8}) and (\ref{eqn:D15})]
\begin{equation}
  \label{eq:el_Phi}
  \frac{\Phi}{V} = \sum_{\beta=x,y,z} \left[ \eta_s \kappa_{\beta\beta}^2 
    + \frac{1}{4} \eta_p \frac{( \dot{c}_{\beta\beta}-2\kappa_{\beta\beta} c_{\beta})^2}{c_{\beta\beta}} \right] 
\end{equation}
where $\eta_p$ is defined by $\eta_p=G\tau$ and represents the polymer 
contribution to the viscosity.
 
The evolution equations of $\tens{c}$-tensor can then be derived by 
$\partial \mathscr{R}/\partial \dot{\tens{c}}=0$,
\begin{equation}
  \dot{c}_{\beta\beta} = 2 \kappa_{\beta\beta} c_{\beta\beta} - \frac{1}{\tau} (c_{\beta\beta} - 1).
  \label{eq:el_cdotbeta}
\end{equation}
We write them in terms of the radial and axial components explicitly,
\begin{eqnarray}
  \dot{c}_{rr} &=& - \dot{\varepsilon} c_{rr} -\frac{1}{\tau} ( c_{rr} -1) ,\\
  \dot{c}_{zz} &=&  2\dot{\varepsilon} c_{zz} -\frac{1}{\tau} ( c_{zz} -1) .
\end{eqnarray}

The stress tensor can also be derived from Rayleighian by differentiating  
$\mathscr{R}$ with respect to the velocity gradient tensor $\tens{\kappa}$,
\begin{equation}
  \sigma_{\beta\beta} = 2 \eta_s \kappa_{\beta\beta} - \eta_p (\dot{c}_{\beta\beta} 
    - 2 \kappa_{\beta\beta} c_{\beta\beta} ) = 2 \eta_s \kappa_{\beta\beta} + G(c_{\beta\beta} -1 ),
\end{equation}
where we have used the result of Eq.~(\ref{eq:el_cdotbeta}). 
In the radial and axial directions, the equations become
\begin{eqnarray}
   \sigma_{rr} &=& - \eta_s \dot{\varepsilon} + G(c_{rr} - 1),\\
   \sigma_{zz} &=& 2 \eta_s \dot{\varepsilon} + G(c_{zz} - 1) .
\end{eqnarray}
These are the constitutive equations for the Oldroyd-B model.

The tensile force is given by $T = \partial \mathscr{R} / \partial \dot{L} = (\partial \mathscr{R} / \partial \dot{\lambda} ) / L_{\rm ini} = ( \partial \mathscr{R} / \partial \dot{\varepsilon} ) /L$,
\begin{equation}
    \frac{T}{\pi R^2} = 3 \eta_s \dot{\varepsilon} + G (c_{zz}-c_{rr}) + G \alpha \sqrt{\lambda}.
\end{equation}
There are three contributions to the tensile force: the solvent viscosity, the polymer stretching, and the surface tension.

%====================================================
\subsection{Onsager principle as an approximation}

% CPBS

We now consider the case that the stretching takes place non-uniformly: the
overall length of the filament is stretched by factor $\lambda$, but the stretching
has taken place non-uniformly; some part has been stretched strongly, other
part weakly.  In this case the local strain of the filament  becomes a function of 
position.   This situation has been studied by several groups 
\cite{Forest1990, Eggers1997, LiJie2003, Clasen2006}.   Assuming 
that the deformation is uniaxial extension, they derived a set of equations
which describes the time evolution of the filament radius $h(z,t)$. 
The same set of equations can also be derived by Onsager principle 
(see Appendix \ref{app:1dEqn}). 

An advantage of using Onsager principle is that we can 
do approximate calculation by utilizing the fact that the 
evolution equations is derived by variational principle
\cite{Doi2015, Doi2016}.  
By assuming certain form for the non-equilibrium state of the system 
which involves small set of parameters, one can derive the time evolution 
of these parameters by the Onsager principle.
This has been used in various problems \cite{Tomari1995, MengFanlong2016a, ManXingkun2016, DiYana2016, XuXianmin2016, 2017_stratification}.

In the following we take the same strategy for the present problem.
To simplify the equation, we assume that principal value of the
$\tens{c}$-tensor is  written as 
\begin{equation}
  c_{zz} = \lambda_p^2, \quad \quad c_{rr} = \frac{1}{\lambda_p}. 
                         \label{eq:D201}
\end{equation}
The reason for this approximation is as follows.
If the relaxation time $\tau$ of the Oldroyd-B model is infinitely large, 
$\tens{c}$-tensor deforms affinely and Eq. (\ref{eq:D201}) is strictly
satisfied due to the incompressible condition. 
If the relaxation time is finite, Eq.~(\ref{eq:D201}) may not correct, but we can assume this form as a possible non-equilibrium state of the system.
For strongly stretched filament, the radial component $c_{rr}$ is order of magnitude less than the axial component $c_{zz}$, therefor it plays an insignificant role in the dynamics.

Substituting Eq.~(\ref{eq:D201}) into Eqs.~(\ref{eq:el_A1}) and (\ref{eq:el_Phi}), 
the free energy and the dissipation function can be reexpressed as
\begin{eqnarray}
  \frac{A}{V} &=& G \left[ \frac{1}{2} \left( \lambda_p^2 + \frac{2}{\lambda_p} \right)
    + 2\alpha \sqrt{\lambda} \right] , 
  \label{eq:AAA} \\
  \frac{\Phi}{V} &=& \frac{3}{2} \eta_s \Big( \frac{\dot{\lambda}}{\lambda} \Big)^2 
    + \eta_p \Big( \lambda_p^2 + \frac{1}{2\lambda_p} \Big) 
    \Big( \frac{\dot{\lambda}_p}{\lambda_p} - \frac{\dot{\lambda}}{\lambda} \Big)^2 .
  \label{eq:Phi}
\end{eqnarray}
The evolution equation is given by $\partial \mathscr{R}/\partial \dot{\lambda}_p=0$,
\begin{equation}
  \label{eq:el_lp}
  \frac{\dot{\lambda}_p}{\lambda_p} - \frac{\dot{\lambda}}{\lambda} = 
    -\frac{1}{2\tau} \frac{ \lambda_p^2 - \frac{1}{\lambda_p}} {\lambda_p^2 + \frac{1}{2\lambda_p}} .
\end{equation}
The tensile force is given by  $T= (\partial \mathscr{R} / \partial \dot{\lambda}) / L_{\rm ini}$,
\begin{equation}
  \label{eq:el_T1}
  \frac{T}{\pi R^2} = 3 \eta_s \dot{\varepsilon} + G (\lambda_p^2 - \frac{1}{\lambda_p} ) + G \alpha \sqrt{\lambda}.
%  \frac{T}{\pi R_{\rm ini}^2} &=& G \left[ \frac{3\eta_s}{\eta_p} \dot{\varepsilon} \tau \frac{1}{\lambda} + (\lambda_p^2 - \frac{1}{\lambda_p} ) \frac{1}{\lambda} + \alpha \lambda^{-1/2} \right].
\end{equation}

%%%%%%%%%%%%%%%%%%%%%%%%%%%%%%%%%%%%%%%%%%%%%
\section{Formation of Beads-on-string} 
%\subsection{Formation of beads-on-string} 

\begin{figure}[htbp]
  \centering
  \includegraphics[width=0.7\textwidth]{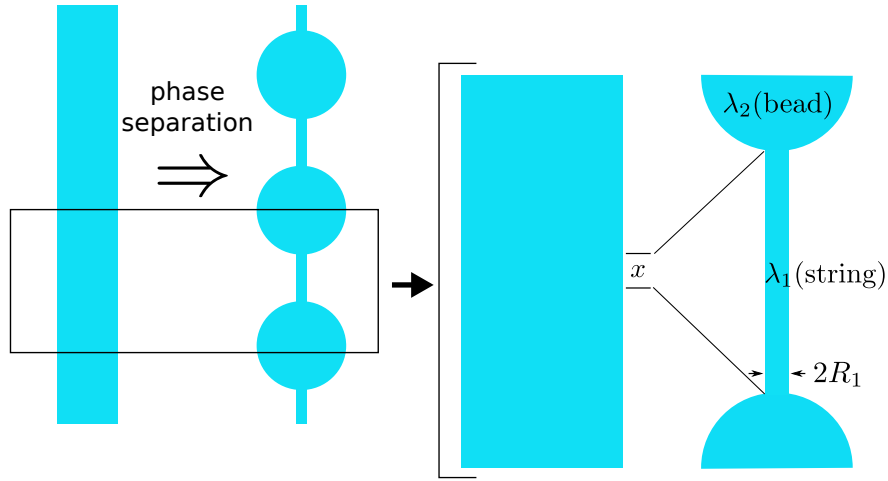}
  \caption{Formation of the beads-on-string structure due to the phase separation. The enlarged part is for one length period. Fraction $x$ of the initial cylinder becomes the string part, while the fraction $(1-x)$ goes to the beads.} 
  \label{fig:sketch2}
\end{figure}

When the filament is stretched and then held fixed at both ends, the cylindrical configuration may become unstable and the bead-on-string configuration appears.
This is shown schematically in Fig.~\ref{fig:sketch2}.

Analysis of this process is not easy as one has to solve the set of evolution equations for the whole profile, as it was done in Ref.~\cite{Clasen2006}.  
However, since the process takes place in a time scale much shorter than the polymer relaxation time $\tau$, we can obtain the state at the end of this process in a simple way. 
We consider the limit that $\tau$ is infinity. 
From Eq.~(\ref{eq:el_lp}), we see in this case the polymer elongation is affine, i.e., $\lambda_p = \lambda$. 
The polymer solution can then be regarded as an elastic gel, and the problem becomes equivalent to calculate the equilibrium state of a stretched gel \cite{Mora2010, XuanChen2016}.  
The free energy (\ref{eq:AAA}) becomes
\begin{equation}
  \label{eq:A_affine}
  \frac{A}{V} = G \left[ \frac{1}{2} \left( \lambda^2 + \frac{2}{\lambda} \right)
    + 2\alpha \sqrt{\lambda} \right] .
\end{equation}
In this quasi-equilibrium state, the tensile force is given by
\begin{equation}
  \label{eq:el_T0}
  T = \frac{ \partial A }{ \partial \lambda } \frac{1}{L_{\rm ini}} = \pi R^2_{\rm ini} G \left[ \lambda - \frac{1}{\lambda^2} + \alpha \lambda^{-1/2} \right].
\end{equation}
This expression is consistent with Eq. (\ref{eq:el_T1}) if one uses the affine condition and neglects the solvent contribution.

\begin{figure}[htbp]
  \centering
  (a)\includegraphics[width=0.6\columnwidth]{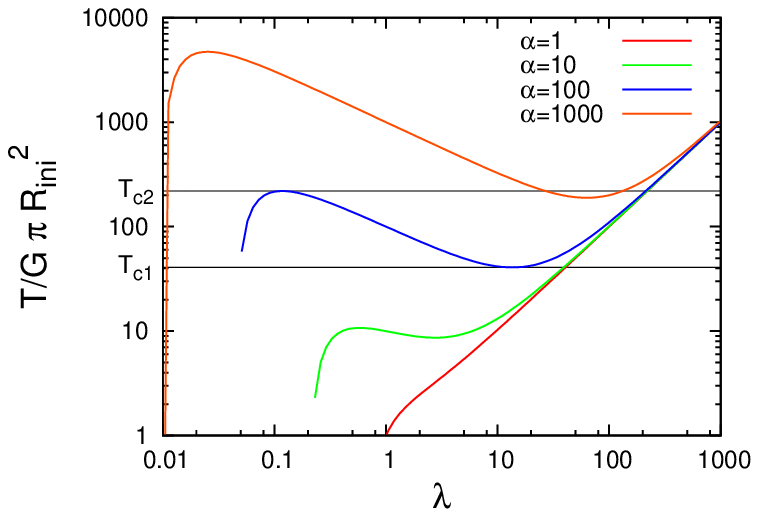}
  (b)\includegraphics[width=0.6\columnwidth]{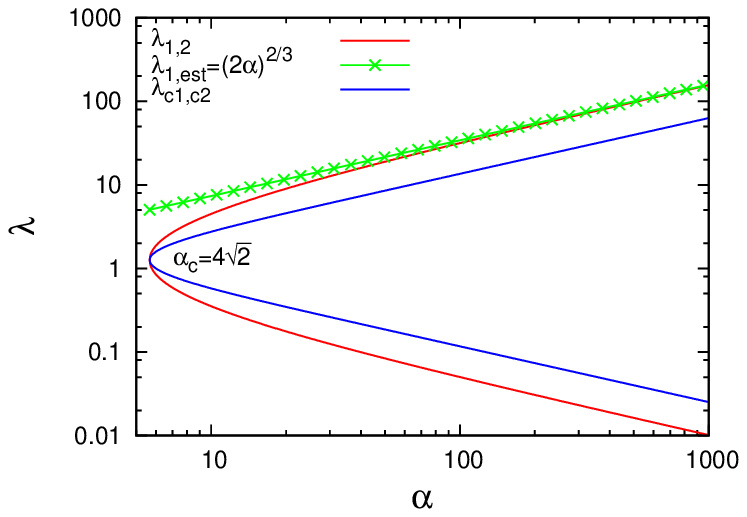}
  \caption{(a) The tensile force as a function of $\lambda$. (b) Phase diagram in the $\alpha$-$\lambda$ coordinates.}
  \label{fig:tension}
\end{figure}

% \begin{figure}[htbp]
%   \centering
%   \includegraphics[width=0.7\columnwidth]{fig2}
%   \caption{The tensile force $T$ as a function of $\lambda$. For $\alpha=100$, the local minimum and maximum correspond to $T_{c1}$ and $T_{c2}$, respectively.}
%   \label{fig:tension}
% \end{figure}

Representative plots of $T$ as a function of $\lambda$ are shown in Fig.~\ref{fig:tension}(a). 
When $\alpha$ is large, in the range of $[T_{c1},T_{c2}]$, there are three solutions for $\lambda$. 
The local extremes are given by 
\begin{equation}
  \frac{1}{\pi R_{\rm ini}^2} \frac{\ud T}{\ud \lambda} = G \left[ 1+ \frac{2}{\lambda^3} - \frac{1}{2} \alpha \lambda^{-3/2} \right] = 0. 
\end{equation}
The above equation has a solution only if 
\begin{equation}
  \alpha \ge 4 \sqrt{2}.
\end{equation}
If the above condition is satisfied, the two solutions are
\begin{equation}
  \lambda_{c1} = \left[ \frac{1}{8} ( \alpha - \sqrt{\alpha^2 - 32} ) \right]^{-2/3} ,
% \simeq  \left( \frac{2}{\alpha} \right)^{-2/3}, \\
  \quad
  \lambda_{c2} = \left[ \frac{1}{8} ( \alpha + \sqrt{\alpha^2 - 32} ) \right]^{-2/3} .
% \simeq  \left( \frac{\alpha}{4} - \frac{2}{\alpha} \right)^{-2/3} .
\end{equation}
The value of $\lambda_{c1}$ and $\lambda_{c2}$ are plotted as a function of $\alpha$ in Fig.~\ref{fig:tension}(b).

The formation of beads-on-string structure is similar to a phase separation process (see Fig.~\ref{fig:sketch2}).
%Initially the filament is uniformly elongated while the tensile force increases.
The uniform cylindrical filament is unstable and phase-separated into two parts: Fraction $x$ has a large elongation $\lambda_1$ and forms the string part;
fraction $(1-x)$ has a small elongation $\lambda_2$.
The small elongation part would eventually relax and form the spherical bead.
However, in the following, we proceed the calculation assuming that the part of small elongation has a cylindrical shape.  
This does not affect the final results since the elongation $\lambda_2$ and the free energy $A(\lambda_2)$ of this part are negligibly small compared with those of the string part, and the final equations become independent of the shape of this part.

To calculate $\lambda_1$ and $\lambda_2$, one need to minimize the total free energy under the constraint 
\begin{equation}
  x \lambda_1 + (1-x) \lambda_2 = \bar{\lambda},
  \label{eq:constraint}
\end{equation}
where $\bar{\lambda}$ is the fixed elongation before phase-separation.
This corresponds to the minimization of the following function
\begin{equation}
  \mathscr{A} = x A(\lambda_1) + (1-x) A(\lambda_2) - \xi \left[ x \lambda_1 + (1-x) \lambda_2 - \bar{\lambda} \right],
  \label{eq:totalA}
\end{equation}
where $\xi$ is a Lagrangian multiplier to enforce the constraint (\ref{eq:constraint}).

Minimization of $\mathscr{A}$ with respect to $\lambda_1$, $\lambda_2$ and $x$ gives
\begin{eqnarray}
  \label{eq:cocond_T1}
  \frac{ \partial \mathscr{A} }{\partial \lambda_1} = 0  & \quad \Rightarrow \quad & 
    \frac{\partial A}{\partial \lambda_1} = \xi = T L_{\rm ini}, \\
  \label{eq:cocond_T2}
  \frac{ \partial \mathscr{A} }{\partial \lambda_2} = 0 & \quad \Rightarrow \quad &
    \frac{\partial A}{\partial \lambda_2} = \xi = T L_{\rm ini}, \\
  \label{eq:cocond_TL}
  \frac{ \partial \mathscr{A} }{\partial x} = 0 & \quad \Rightarrow \quad & 
    A(\lambda_1) - A(\lambda_2) = \xi (\lambda_1 - \lambda_2) 
    = T L_{\rm ini} (\lambda_1 - \lambda_2) .
\end{eqnarray}

In Eqs.~(\ref{eq:cocond_T1}) and (\ref{eq:cocond_T2}), we have written the Lagrangian multiplier as $TL_{\rm ini}$ according to Eq.~(\ref{eq:el_T0}).
These two equations implies the force balance between the bead and the string.
Equation (\ref{eq:cocond_TL}) gives the coexistence condition of the bead and the string. 

Since the polymers are relaxed in the bead part, one can neglect the free energy contribution from the bead part, then Eq.~(\ref{eq:cocond_TL}) gives
\begin{equation}
  A(\lambda_1) = T L_{\rm ini} (\lambda_1 - \lambda_2) \simeq T L_{\rm ini} \lambda_1 .
\end{equation}
From the free energy expression (\ref{eq:A_affine}), the above equation becomes
\begin{equation}
%  &&\frac{\lambda_1^2}{2} + \frac{1}{\lambda_1} + 2 \alpha \lambda_1^{1/2} = \frac{T}{\pi R_{\rm ini}^2 G} \lambda_1, \\
  \frac{\lambda_1^2}{2} + 2 \alpha \lambda_1^{1/2} = \frac{T}{\pi R_{\rm ini}^2 G} \lambda_1,
  \label{eq:co_cond}
\end{equation}
where we have neglected a term $1/\lambda_1$ which is much smaller than other terms.
% Using the expression for $\alpha$ (\ref{eq:alpha}) and $R_1$ (\ref{eq:LR}), one can simplify Eq.~(\ref{eq:co_cond}) to
% \begin{equation}
%   \label{eq:co_cond2}
%   \frac{G R_{\rm ini}^4}{2 R_1^4} + \frac{2\gamma}{R_1} = \frac{T}{\pi R_1^2}.
% \end{equation}
% This is the same equation~(3.10) derived in Ref. \cite{Clasen2006} if one (1) neglect the small contribution due to $1/\bar{R}$ term; and (2) makes the same nondimensional transformation $\gamma=1$ and $R_{\rm ini}=1$. 
This coexistent condition is consistent with the result of Ref.~\cite{Clasen2006},
where the condition is derived by matching the string to an almost spherical bead.

The value of $\lambda_1$ at coexistence can be calculated by Eqs.~(\ref{eq:el_T0}) and (\ref{eq:co_cond}) (again neglecting the smallest term $1/\lambda^2$). 
The results are
\begin{equation}
  \lambda_1 = (2\alpha)^{2/3}, \quad
  R_1 = R_{\rm ini} (2\alpha)^{-1/3} = R_{\rm ini}^{4/3} \left( \frac{G}{2\gamma} \right)^{1/3}. 
  \label{eq:R1}
\end{equation} 
One can compare the exact solution from Eqs.~(\ref{eq:cocond_T1})--(\ref{eq:cocond_TL}) with the approximated solution of Eq.~(\ref{eq:R1}). 
This is shown in Fig.~\ref{fig:tension}(b).
The approximation works very good when the elasto-capillary number is large, i.e., $\alpha > 100$.  
In Ref.~\cite{Anna2001}, they reported experiments using Boger fluids with $\alpha=30,126,400$.

Equation (\ref{eq:R1}) also gives the string size right after the bead-on-string structure forms.
The scaling $(G/2\gamma)^{1/3}$ again agrees with Ref.~\cite{Clasen2006}.
Similar energy argument was used to interpret the Plateau-Rayleigh instability of an elastic cylinder \cite{XuanChen2017}.

%%%%%%%%%%%%%%%%%%%%%%%%%%%%%%%%%%%%%%%%%%%%%
\section{Elasto-capillary Thinning}

%===========================================
\subsection{Dynamical coexistence condition}

\begin{figure}[htbp]
  \centering
  \includegraphics[width=0.4\textwidth]{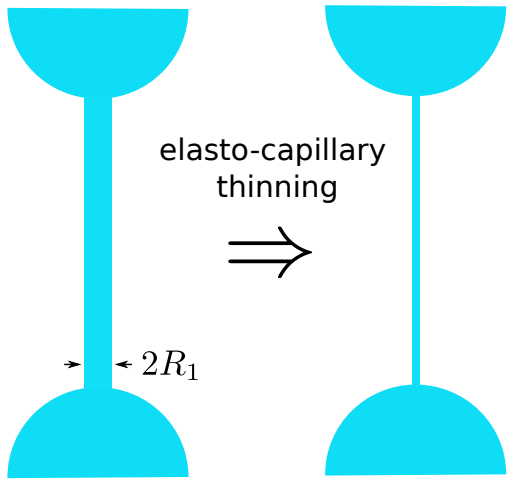}
  \caption{The string shrinks with time due to the elasto-capillary effect. Only one length period of the beads-on-string structure is shown here.} 
  \label{fig:sketch3}
\end{figure}

After beads-on-string structure forms, the string part continues thinning under surface tension (see Fig.~\ref{fig:sketch3}). 
From phase separation point-of-view, this corresponds to the dynamical coexistence between the bead and the string.
In contrast to equilibrium situation, where the coexistence is obtained by minimize the total free energy (\ref{eq:totalA}), we shall derive the dynamical coexistent condition by minimizing the total Rayleighian.
%We assume fraction $x$ remains in the string, while fraction $1-x$ has gone to the beads.

Free energy and dissipation function of the whole system are given by
\begin{eqnarray}
  A_{\rm tot} &=& x A(1) + (1-x) A(2) ,     \label{eqn:Atot}\\    
  \Phi_{\rm tot} &=& x \Phi(1) + (1-x) \Phi(2) ,   \label{eqn:Phitot}
\end{eqnarray}
where $A(i)=A(\lambda_{i}, \lambda_{pi})$ and $\Phi(i)=\Phi(\dot{\lambda}_i, \dot{\lambda}_{pi}, \lambda_i, \lambda_{pi})$ are given by Eqs.~(\ref{eq:AAA}) and (\ref{eq:Phi}), respectively.
The label 1 corresponds to the string part, while label 2 for the bead part.

In writing Eqs.~(\ref{eqn:Atot}) and (\ref{eqn:Phitot}), we have ignored the terms associated with the transition region between beads and string. 
This approximation is justified for the free energy part, but not so for the dissipation function part.  
Here we shall proceed assuming Eq.~(\ref{eqn:Phitot}), and come back to this problem later.

The Rayleighian of the whole system is written as
\begin{equation}
  \label{eq:tot_R}
  \mathscr{R} = \Phi_{\rm tot} + \dot{A}_{\rm tot} - \xi \left[ \dot{x} (\lambda_1 - \lambda_2) + x \dot{\lambda}_{1} + (1-x) \dot{\lambda}_{2} - \dot{\bar{\lambda}} \right],
\end{equation}
where the $\xi$ term is a Lagrangian multiplier to enforce the time derivative of constraint (\ref{eq:constraint}).

The evolution equations are obtained by 
\begin{eqnarray}
  \label{eq:cocond_lami}
  \frac{ \partial \mathscr{R} }{ \partial \dot{\lambda}_i } =0 
    &\quad \Rightarrow \quad& 
    \frac{\partial \Phi(i)}{\partial \dot{\lambda}_i} + \frac{\partial A(i)}{\partial \lambda_i} = \xi = T L_{\rm ini}, \\
  \label{eq:cocond_lampi}
  \frac{ \partial \mathscr{R} }{ \partial \dot{\lambda}_{pi} } =0 
    &\quad \Rightarrow \quad &
    \frac{\partial \Phi(i)}{\partial \dot{\lambda}_{pi}} + \frac{\partial A(i)}{\partial \lambda_{pi}} = 0, \\
  \label{eq:cocond_xdot}
  \frac{ \partial \mathscr{R} }{ \partial \dot{x} }=0 
    &\quad \Rightarrow \quad &
    A(1) - A(2) = \xi (\lambda_1 - \lambda_2) = T L_{\rm ini} (\lambda_1 - \lambda_2).
\end{eqnarray}
%These three equations have clear physical meaning. 
Equation (\ref{eq:cocond_lami}) is the force balance equation, where the tensile force is given by $T = (\partial \mathscr{R} / \partial \dot{\lambda}) / L_{\rm ini}$. 
The Lagrangian multiplier can also be written in term of the true stress tensor, $\zeta = T L_{\rm ini} = \sigma_{T} V$.
Equation (\ref{eq:cocond_lampi}) results in the evolution equation for polymer conformation $\lambda_{pi}$, i.e., the constitutive equations for the Oldroyd-B model.

The last equation (\ref{eq:cocond_xdot}) is the main result of our paper.
In contrast to the previous two equations, there is no existing equation of motion corresponding to Eq. (\ref{eq:cocond_xdot}).
This equation is obtained by minimizing $\mathscr{R}$ with respect to $\dot{x}$. 
This resembles the equality of chemical potentials for equilibrium, which is to minimize the total free energy with respect to the fraction $x$. 
Therefore, one may view it as a dynamical coexistence condition between the bead and the string. 
In a more general setting, the dissipation function could include terms like $\dot{x}^2$, which is associated with the dissipation in the transition region. 
We have neglected those contributions. 
Therefore, the dynamical coexistence condition (\ref{eq:cocond_xdot}) takes the same form as in equilibrium (\ref{eq:cocond_TL}), though the tensile force now is given by $T = (\partial \mathscr{R} / \partial \dot{\lambda}) / L_{\rm ini}$ instead of $T = (\partial A / \partial \lambda ) / L_{\rm ini}$.

%==================================
\subsection{Exponential thinning of the string}

The polymers in the beads are relaxed so that $A(1) \gg A(2)$. 
The string has a much larger elongation than the bead, $\lambda_1 \gg \lambda_2$. 
Using these relations, Eq.~(\ref{eq:cocond_xdot}) is simplified to
\begin{equation}
  \label{eq:sigmaEV}
  T L_{\rm ini} = \frac{A(1)}{\lambda_1}=\frac{ A(\lambda_1,\lambda_{p1}) }{ \lambda_1 }.
\end{equation}
We only need to consider the string portion.
The evolution equations for the string are
\begin{eqnarray}
  \label{eq:evol_lam}
  \frac{\partial \Phi(1)}{\partial \dot{\lambda}_1} + \frac{\partial A(1)}{\partial \lambda_1} 
    &=& \frac{ A(1) }{ \lambda_1 } , \\
  \label{eq:evol_lamp}
  \frac{\partial \Phi(1)}{\partial \dot{\lambda}_{p1}} + \frac{\partial A(1)}{\partial \lambda_{p1}} &=& 0.
\end{eqnarray}

We shall drop the label 1 in the subscript for simplicity.
The evolution equations can be obtained by using Eqs.~(\ref{eq:AAA}) and (\ref{eq:Phi}),
\begin{eqnarray}
  \label{eq:lambda_ob1}
  \frac{\dot{\lambda}_p}{\lambda_p} - \frac{\dot{\lambda}}{\lambda} 
    - \frac{3\eta_s}{2\eta_p} \frac{1}{\lambda_p^2+\frac{1}{2\lambda_p}} 
      \frac{\dot{\lambda}}{\lambda} 
     &=& - \frac{1}{4\tau} \frac{ \lambda_p^2+\frac{2}{\lambda_p}+ 2\alpha \sqrt{\lambda}}
      {\lambda_p^2+\frac{1}{2\lambda_p}} , \\
  \label{eq:lambda_ob2}
   \frac{\dot{\lambda}_p}{\lambda_p} - \frac{\dot{\lambda}}{\lambda} 
     &=& - \frac{1}{2\tau}
        \frac{\lambda_p^2 - \frac{1}{\lambda_p}}{\lambda_p^2+\frac{1}{2\lambda_p}} .
\end{eqnarray}

We consider the late stage when polymers are strongly stretched $\lambda_p \gg 1$, so $1/\lambda_p$ terms can be neglected.
The evolution equations are simplified as  
\begin{eqnarray}
  \label{eq:OB_ev1}
  \frac{\dot{\lambda}_p}{\lambda_p} - \frac{\dot{\lambda}}{\lambda} 
     &=& - \frac{1}{4\tau} \frac{ \lambda_p^2 +2 \alpha \sqrt{\lambda}}
      {\lambda_p^2} , \\
  \label{eq:OB_ev2}
  \frac{\dot{\lambda}_p}{\lambda_p} - \frac{\dot{\lambda}}{\lambda} 
     &=& - \frac{1}{2\tau} .
\end{eqnarray}
We immediately obtain the following relation between $\lambda$ and $\lambda_p$
\begin{equation}
  \label{eq:OB_lambda}
  \lambda_p^2 = 2\alpha \sqrt{\lambda}.
\end{equation}
Equations (\ref{eq:OB_ev1}) and (\ref{eq:OB_ev2}) can then be solved as
\begin{eqnarray}
  \label{eq:OB_l}
  \lambda &=& \lambda_{\rm e} \exp \left[ \frac{2}{3\tau} (t-t_{\rm e}) \right], \\
  \label{eq:OB_lp}
  \lambda_p^2 &=& 2 \alpha \lambda_{\rm e}^{1/2} \exp \left[ \frac{1}{3\tau} (t-t_{\rm e}) \right] .
\end{eqnarray}
The above equations indicates that both $\lambda$ and $\lambda_p$ increase exponentially.
The quantities $(t_{\rm e}, \lambda_{\rm e})$ are values at the start of the exponential behavior.
Equation~(\ref{eq:OB_l}) implies that the string radius decreases exponentially with a time constant $3\tau$, 
\begin{equation}
  R(t) = R_{\rm e} \exp \left[ - \frac{1}{3\tau} (t-t_{\rm e}) \right].
\end{equation}

% The tensile stress can be calculated
% by considering the force balance in the axial and radial directions
% \begin{equation}
%   \frac{T}{\pi R^2} = ( \sigma_{zz} - p ) + \frac{2\gamma}{R} , \quad
%   \sigma_{xx}-p = - \frac{\gamma}{R} .
% \end{equation}
% Eliminating the pressure and using the constitutive equation (\ref{eq:el_sigma}),
The tensile force is given by Eq.~(\ref{eq:el_T1})
\begin{equation}
  \label{eq:OB_tension0}
  \frac{T}{\pi R^2} = 3\eta_s \dot{\varepsilon} + G (\lambda_p^2 - \frac{1}{\lambda_p} ) + \frac{\gamma}{R},
\end{equation}
There are three contributions to the tensile force: The first is from the solvent viscosity, which in general is much smaller than the other two terms and can be neglected. 
The second is from the polymer elasticity, and the last one comes from the surface tension.

Writing the last two terms in (\ref{eq:OB_tension0}) explicitly (while neglecting the small term of $1/\lambda_p$)
\begin{eqnarray}
%  3\eta_s \dot{\varepsilon} &=& 2G \frac{\eta_s}{\eta_p} , \\
  \label{eq:OB_Gczz}
  G \lambda_p^2 &=&  2 G \alpha \lambda_1^{1/2} \exp \left[ \frac{1}{3\tau} (t-t_{\rm e}) \right], \\
  \frac{\gamma}{R} &=& G\alpha \lambda_1^{1/2} \exp \left[ \frac{1}{3\tau} (t-t_{\rm e}) \right], 
\end{eqnarray}
we obtain the tensile stress
\begin{equation}
  \label{eq:OB_tension}
  \frac{T}{\pi R^2} \simeq 3 G \alpha \lambda_1^{1/2} \exp \left[ \frac{1}{3\tau} (t-t_{\rm e}) \right] = \frac{3\gamma}{R}.
\end{equation}

%===========================================
\subsection{Comparison to previous results}

We now compare our theory with the classical 
theory of Entov and Hinch \cite{Entov1997}.
They considered the time evolution of a uniformly stretched viscoelastic string by setting up the force balance equations in axial ($z$) and radial ($r$) direction. 
\begin{eqnarray}
  \frac{T}{\pi R^2} &=& \sigma_{zz} - p + \frac{2 \gamma}{R} ,
                       \label{eq:EH1} \\
  0 &=& \sigma_{rr} - p + \frac{\gamma}{R} .
                       \label{eq:EH2} 
\end{eqnarray}
where $p$ is the pressure in the solution, and $\sigma_{zz}$ and $\sigma_{rr}$ are components of the stress tensor, which are related to $c_{zz}$ and $c_{rr}$ by the constitutive equations. The above two equations include 5 unknowns, $T$, $p$, $c_{zz}$, $c_{rr}$ and $R$. 
This is supplemented by two evolution equations for $c_{zz}$ and $c_{rr}$.  
Hence the number of equations is 4, but the number of unknowns is 5.
One equation is missing.

To close the equation, Entov and Hinch assumed an extra equation
$\sigma_{zz} - p=0$, and derived the following results
\begin{eqnarray}
  R(t) &=& R_{\rm e} \exp \left[ - \frac{1}{3\tau} (t-t_{\rm e}) \right] ,
                       \label{eq:EH3} \\
  \frac{T}{\pi R^2} &=& \frac{2 \gamma}{R}.
                       \label{eq:EH4}
\end{eqnarray}
Since the exponential behavior of Eq.~(\ref{eq:EH3}) was found to agree well with 
experiments, the theory has been regarded as a successful classical theory for the beads-on-string phenomena.

On the other hand, their assumption $\sigma_{zz} - p=0$ has been questioned for its validity. 
Stelter et al. \cite{Stelter2000} solved the problem using a different condition $\sigma_{zz}-p = \gamma/R$. 
They obtained the same exponential decay for $R(t)$ , but different numerical coefficient for the tension $T/\pi R^2$.
In fact, one can show that any assumption of the form $\sigma_{zz}-p = \chi \, \gamma /R$ ($\chi$ being a certain numerical constant) gives the same exponential decay for the string radius. 
The polymer extension, on the other hand, will depend on $\chi$,
\begin{equation}
  \label{eq:chi}
  \lambda_p = \sqrt{1+\chi} \alpha^{1/2} \lambda_{\rm e}^{1/4} \exp \left[ \frac{1}{6\tau} ( t - t_{\rm e}) \right].
\end{equation}

Here we derived the missing equation using the Onsager principle. 
The physics behind our equation (\ref{eq:cocond_xdot}) is the a dynamical coexistence condition between the string and the bead. 
This condition is similar to the phase coexistent condition in equilibrium.  
However, it must be noted that we are dealing with non-equilibrium process, and that the proper quantity to be minimized is the total Rayleighian (\ref{eq:tot_R}).  

Clasen et al. \cite{Clasen2006} conducted a numerical analysis for the time evolution of the entire filament including both the bead and the string.  
Using the numerical results, they showed that the tensile force is given by Eq.~(\ref{eq:OB_tension}), not by Eq.~(\ref{eq:EH4}).  
Therefore, we believe that the coexistence condition (\ref{eq:cocond_xdot}) we have used to derive Eq.~(\ref{eq:OB_tension}) is the equation which should replace the controversial equation of Entov and Hinch.   
The immediate consequence is that the polymer chains are more stretched in the thinning process, as the front factor in Eq.~(\ref{eq:chi}) is equal to $\sqrt{2}$, instead of $1$ from Entov and Hinch.

\begin{figure}[htbp]
  \centering
  \includegraphics[width=0.8\textwidth]{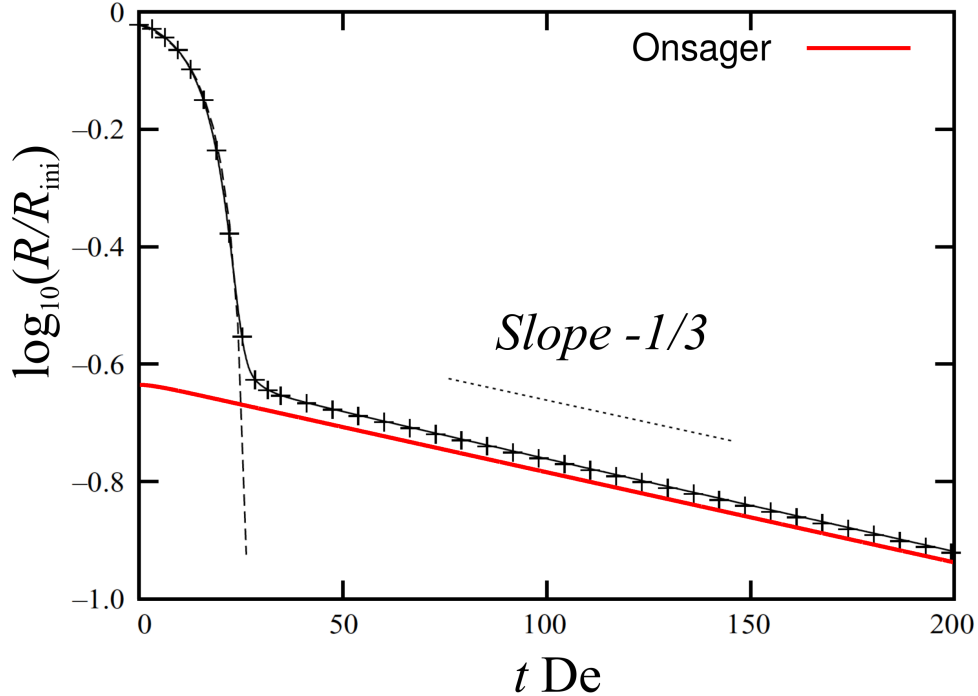}
  \caption{Comparison between the results from Onsager principle [Eqs.~(\ref{eq:lambda_ob1}) and (\ref{eq:lambda_ob2})] and the numerical results of Clasen \cite{Clasen2006}. The parameters are $\alpha=40$ and $\eta_p/\eta_s=3.0$. In Ref. \cite{Clasen2006}, the unit of the time is the capillary time $\tau_{\gamma} = ( \rho a^3 / \gamma)^{1/2}$, while we used the polymer's relaxation time $\tau$. The difference in the time units leads to the appearance of Deborah number (${\rm De}=\tau/\tau_{\gamma}=94.9$) in the x-axis.}
  \label{fig:Clasen}
\end{figure}

In Fig. \ref{fig:Clasen}, we show a comparison between our results and numerical results of Clasen et al.\cite{Clasen2006} 
Our results are obtained by numerically solving the evolution equations (\ref{eq:lambda_ob1}) and (\ref{eq:lambda_ob2}). 
The parameters are $\alpha=40$ and $\eta_p/\eta_s=3.0$. 
Since in our framework, the initial phase-separation happens over a very short time, our results only give the evolution which starts at $\lambda(t=0)=\lambda_p(t=0)=(2\alpha)^{2/3}$.

%%%%%%%%%%%%%%%%%%%%%%%%%%%%%%%%%%%%%%%%%%%%%%%%%%%%%%
\section{Conclusion}

In conclusion, we have developed a framework to analyze the viscoelastic fluid based on Onsager principle, and showed explicitly how to apply this framework to the beads-on-string structure formation in a viscoelastic filament. 
Using the Oldroyd-B model for the polymer chain, we derived evolution equations for the viscoelastic filaments. 
Without using any assumptions on the axial stress, we were able to derive the exponential thinning based on the coexistence condition (\ref{eq:cocond_xdot}).  
Most of our results agree with the classical theory of Entov and Hinch \cite{Entov1997}, but there are differences: 
\begin{itemize}
\item[(i)] the tensile force acting on the string [Eqs.~(\ref{eq:OB_tension}) and (\ref{eq:EH4})]. 
\item[(ii)] the elongation of the polymer chains during the exponential thinning [Eqs.~(\ref{eq:chi})]. 
\end{itemize}
In these differences, our results agree with the numerical solution of Clasen et al. \cite{Clasen2006}.

% The validity of the coexistence condition we have derived in this paper relies on whether the expression of Rayleighian (\ref{eq:tot_R}) is valid or not. 
% The Rayleighian we have used assumes that both energy dissipation function and the free energy are given by the sum of the bulk part, and ignore any contribution arising from the transition region.
% At the transition region, the fluid in the string moves into the bead part, and the stretched polymer chains relax to equilibrium. 
% Therefore there is an extra dissipation which we may write symbolically as $\Phi_{\rm tr}(\dot{x})$. 
% In the beads-on-string structure, the term $\Phi_{\rm tr}(\dot{x})$ is negligible compared with the string term $\Phi(1)$.
% This justifies our approach, and the final result is the dynamical coexistent condition (\ref{eq:cocond_xdot}).
% The generalization of current framework to the cases where the term $\Phi_{\rm tr}(\dot{x})$ cannot be neglected is our ongoing work.

In Ref. \cite{Clasen2006}, the derivation of the coexistence condition requires integrating the surface tension through the transition region between the string and the bead. 
Here we have used the Onsager principle to arrive the same condition, and the derivation seems to be simpler. 
The coexistence condition (\ref{eq:cocond_xdot}) we have used in this paper is derived from the expression of Rayleighian (\ref{eq:tot_R}).  
This Rayleighian consists of the string part (which is assumed to be uniform) and the bead part (which is assumed to be in equilibrium), and we ignore the contribution arising from the transition region between the string part and the bead part.  
If the volume of the transition region is finite and does not increase with time, ignoring the contribution of the transition region is justified because the volume of the string part is proportional to the length and can be very large.  
In the present problem, the fluid flows from string to bead and quickly relaxes to equilibrium in the bead. 
In such a situation, the transition region will not grow in time, and the coexistence condition we have used is justified.
However, in other situations, the transition region can grow in time.  
For example, if the distance between the beads is increased quickly, the fluid will flow from
bead to string. 
The entering fluid cannot adjust itself to the string part quickly, and form a transition region.  
In such a case, the volume of the transition region grows in time, and we cannot ignore the contribution of this part.
Therefore the coexistence condition (\ref{eq:cocond_xdot}) is not generally valid:
it is justified only when the Rayleighian associated with the transition region can be ignored.

In the present problem, we may estimate the Rayleighian associated with the 
transition region \cite{Bazilevskii2014, Bazilevskii2015}, and extend the present work to the cases where the dissipation in the transition region cannot be neglected. This is our ongoing work.

%%%%%%%%%%%%%%%%%%%%%%%%%%%%%%%%%%%%%%%%%%%%%%%%%%%%%%
\begin{acknowledgments}
This work was supported by the National Natural Science Foundation of China (NSFC) through the Grant Nos. 21504004 and 21774004. 
M.D. acknowledges the financial support of the Chinese Central Government in the Thousand Talents Program. 
\end{acknowledgments}

%%%%%%%%%%%%%%%%%%%%%%%%%%%%%%%%%%%%%%%%%%%%%%%%%%%%%%
%%%%%%%%%%%%%%%%%%%%%%%%%%%%%%%%%%%%%%%%%%%%%%%%%%%%%%
\appendix
%%%%%%%%%%%%%%%%%%%%%%%%%%%%%%%%%%%%%%%%%%%%%%%%%%%%%%

%%%%%%%%%%%%%%%%%%%%%%%%%%%%%%%%%%%%%%%%%%%%%%%%%%%%%%
\section{Onsager principle on Oldroyd-B model}
\label{app:details}

We consider the viscoelasticity of polymer fluids when the entanglement effect is not important. 
A polymer chain is represented by a dumbbell consisting of two segments connected by a spring with the spring constant $k$.
Let $\mathbf{r}_1$ and $\mathbf{r}_2$ be the position vectors of the two segments. 
The potential energy of the dumbbell is written as 
\begin{equation}
  U(\mathbf{r}_1, \mathbf{r}_2) = \frac{k}{2} (\mathbf{r}_1 - \mathbf{r}_2)^2.
\end{equation}

We assume that the medium surrounding the segments is a Newtonian fluid. 
Suppose the fluid is flowing with velocity gradient tensor $\tens{\kappa}$. 
The flow field is then given by $\tens{\kappa}(t) \cdot \mathbf{r}$.

The conformation of the dumbbell is specified by the distribution function $\Psi(\mathbf{r}_1,\mathbf{r}_2,t)$. We shall derive the time evolution equation for the distribution function $\Psi$ using Onsager principle.

When a segment moves in a flow field, there is a contribution to the energy dissipation due to relative motion between the segment and the surrounding fluid. 
Most simply, this can be written as $\zeta(\dot{\mathbf{r}}_i - \tens{\kappa} \cdot \mathbf{r}_i)^2$, where $\zeta$ is the friction constant, and $(\dot{\mathbf{r}}_i - \tens{\kappa} \cdot \mathbf{r}_i)$ is the relative motion of the segment with respect to the surrounding fluid. 
Hence the energy dissipation per volume is given by
\begin{equation}
  \label{eq:Phi1}
  \Phi = \frac{n_p}{2} \int \ud \mathbf{r}_1 \ud \mathbf{r}_2 
    \left[ \zeta(\dot{\mathbf{r}}_1 - \tens{\kappa} \cdot \mathbf{r}_1)^2
    + \zeta(\dot{\mathbf{r}}_2 - \tens{\kappa} \cdot \mathbf{r}_2)^2 \right]\Psi
\end{equation}
where $n_p$ is the number density of the dumbbell.

It is more convenient to express $\mathbf{r}_1$ and $\mathbf{r}_2$ in terms of the position of the center of mass $\mathbf{R}=(\mathbf{r}_1+\mathbf{r}_2)/2$ and the end-to-end vector $\mathbf{r} = \mathbf{r}_2 - \mathbf{r}_1$ as
\begin{equation}
  \mathbf{r}_1 = \mathbf{R} - \frac{1}{2} \mathbf{r}, \quad 
  \mathbf{r}_2 = \mathbf{R} + \frac{1}{2} \mathbf{r}.
\end{equation}
Equation~(\ref{eq:Phi1}) is then written as
\begin{equation}
  \label{eq:Phi2}
  \Phi = \frac{n_p}{2} \int \ud \mathbf{R} \ud \mathbf{r} 
    \left[ 2\zeta(\dot{\mathbf{R}} - \tens{\kappa} \cdot \mathbf{R})^2
    + \frac{1}{2} \zeta (\dot{\mathbf{r}} - \tens{\kappa} \cdot \mathbf{r})^2 \right]\Psi .
\end{equation}
Since the dumbbells are homogeneously distributed in the solution, we may simply write $\Psi(\mathbf{R},\mathbf{r},t)$ as $\psi(\mathbf{r},t)$ and neglect the motion of the center of mass.
The energy dissipation function is then written as
\begin{equation}
  \label{eq:Phi3}
  \Phi = \frac{n_p}{2} \int \ud \mathbf{r} \,
    \frac{1}{2} \zeta (\dot{\mathbf{r}} - \tens{\kappa} \cdot \mathbf{r})^2 \psi .
\end{equation}

The free energy and its time derivative are given by
\begin{eqnarray}
  A &=& n_p \int \ud \mathbf{r} \left[ k_BT \psi \ln\psi + U \psi \right], \\
  \label{eq:Adot1}
  \dot{A} &=& n_p \int \ud \mathbf{r} \left[ k_BT (\ln\psi + 1) + U \right] \dot{\psi}.
\end{eqnarray}
Using the relation
\begin{equation}
  \label{eq:dotpsi1}
  \dot{\psi} = \frac{\partial \psi}{\partial t} = 
  - \frac{\partial}{\partial \mathbf{r}} \cdot ( \dot{\mathbf{r}} \psi ),
\end{equation}
and integration by parts, Eq.~(\ref{eq:Adot1}) is written as
\begin{equation}
  \dot{A} = n_p \int \ud \mathbf{r} \, \dot{\mathbf{r}} \, \psi 
    \frac{\partial }{\partial \mathbf{r}} (k_BT \ln\psi + U).
\end{equation}
Minimizing $\mathscr{R} = \Phi + \dot{A}$ with respect to $\dot{\mathbf{r}}$, we have 
\begin{equation}
  \frac{1}{2} \zeta ( \dot{\mathbf{r}} - \tens{\kappa} \cdot \mathbf{r} )
    + \frac{\partial}{\partial \mathbf{r}}  (k_BT \ln\psi + U) = 0,
\end{equation}
\begin{equation}
  \dot{\mathbf{r}} = - \frac{2}{\zeta} \left( k_BT \frac{\partial \ln\psi}
    {\partial \mathbf{r}} + \frac{\partial U}{\partial \mathbf{r}} 
    \right) + \tens{\kappa} \cdot \mathbf{r}.
\end{equation}
Therefore the time evolution equation (\ref{eq:dotpsi1}) for $\psi(\mathbf{r},t)$ becomes
\begin{equation}
  \label{eq:dotpsi2}
  \frac{\partial \psi}{\partial t} = \frac{\partial}{\partial \mathbf{r}} 
    \cdot \left( \frac{2k_BT}{\zeta} \frac{\partial \psi}{\partial \mathbf{r}}
    + \frac{2k}{\zeta} \mathbf{r} \psi 
    - \tens{\kappa} \cdot \mathbf{r} \psi \right).
\end{equation}

Let $c_{\alpha\beta} = \frac{k}{k_BT} \langle r_{\alpha} r_{\beta} \rangle $, Eq.~(\ref{eq:dotpsi2}) results in the evolution equation for $\tens{c}$
\begin{equation}
  \dot{\tens{c}} - \tens{\kappa} \cdot \tens{c}  - \tens{c} \cdot \tens{\kappa}^t = - \frac{1}{\tau} ( \tens{c} - \tens{I} ),
\end{equation}
where  $\tau = \zeta/4k$ is the relaxation time.

The polymer contribution to the stress tensor is given by $\tens{\sigma}^{(p)} = \partial \mathscr{R}/ \partial \tens{\kappa}$,
\begin{eqnarray}
  \sigma_{\alpha\beta}^{(p)} &=& - n_p \zeta \int \ud \mathbf{r} ( \dot{r}_{\alpha} - \kappa_{\alpha\mu} r_{\mu} ) r_{\beta} \psi \nonumber \\
    &=& n_p \int \ud \mathbf{r} \, \left[ k_B T \frac{\partial \psi}{\partial r_{\alpha}} r_{\beta} + k r_{\alpha} r_{\beta} \psi \right] \nonumber \\
    &=& - n_p k_BT \delta_{\alpha\beta} + n_p k \langle r_{\alpha}r_{\beta} \rangle .
  \label{eq:sigma1}
\end{eqnarray}
This gives the constitutive equation of Oldroyd-B model
\begin{equation}
  \tens{\sigma}^{(p)} = G ( \tens{c} - \tens{I} ),
\end{equation}
where $G=n_pk_BT$ is the shear modulus.

%=========================================
\subsection{Direct derivation of the c-tensor form}

Assuming the solution of the diffusion equation (\ref{eq:dotpsi2}) is 
\begin{equation}
  \psi(\mathbf{r},t) = (2\pi)^{-3/2} \left[ \det \tens{c}(t) \right]^{-1/2} \exp \left( - \frac{1}{2} \tens{c}^{-1} :\mathbf{r}\mathbf{r} \right)
\end{equation}
The free energy is given by 
\begin{equation}
  A = n_p \int \ud \mathbf{r} ( k_BT \psi \ln\psi + \psi U) = n_p \langle k_BT \ln\psi + U \rangle 
    = \frac{1}{2} n_p k_B T [ {\rm Tr} \tens{c} - \ln \det( \tens{c} ) ]
\end{equation}
We rescale $\tens{c}$ such that at equilibrium $\tens{c}=\tens{I}$,
\begin{eqnarray}
  \label{eq:A3}
  A &=& \frac{1}{2} G \left[ {\rm Tr}(\tens{c}) - \ln \det( \tens{c} ) \right], \\
  \label{eq:Adot3}
  \dot{A} &=& \frac{1}{2} G {\rm Tr} [ (\tens{I} - \tens{c}^{-1}) \cdot \dot{\tens{c}} ].
\end{eqnarray}

The energy dissipation is given by the minimum of 
\begin{equation}
  \label{eq:Phi4}
  \Phi = n_p \frac{\zeta}{2} \langle ( \mathbf{v} - \tens{\kappa} \cdot \mathbf{r} )^2 \rangle,
\end{equation}
under the constraint 
\begin{equation}
  \label{eq:cdot}
  \dot{\tens{c}} = \frac{k}{k_BT} \langle \mathbf{v}\mathbf{r} + \mathbf{r}\mathbf{v} \rangle .
\end{equation} 
We obtain $\mathbf{v}$ by assuming 
\begin{equation}
  \label{eq:V}
  \mathbf{v} (\mathbf{r}) = \tens{V} \cdot \mathbf{r},
\end{equation}
where $\tens{V}$ is a tensor.
Substitute above expression in Eq.~(\ref{eq:Phi4}), 
\begin{equation}
  \label{eq:Phi5}
  \Phi = n_p \frac{\zeta}{2} \langle \mathbf{r} \cdot ( \tens{V}^t - \tens{\kappa}^t) 
         \cdot ( \tens{V} - \tens{\kappa}) \cdot \mathbf{r} \rangle 
    = G \tau {\rm Tr} [ \tens{c} \cdot ( \tens{V}^t - \tens{\kappa}^t) 
         \cdot ( \tens{V} - \tens{\kappa})] .
\end{equation}

From Eqs.~(\ref{eq:cdot}) and (\ref{eq:V}), 
\begin{equation}
  \label{eq:cdotV}
  \dot{\tens{c}} = \tens{V} \cdot \tens{c} + \tens{c} \cdot \tens{V}^t.
\end{equation}
The free energy (\ref{eq:Adot3}) becomes
\begin{equation}
  \dot{A}= \frac{G}{2} {\rm Tr} [ (\tens{I} - \tens{c}^{-1}) \cdot ( \tens{V} \cdot \tens{c} 
      + \tens{c} \cdot \tens{V}^t )] 
    = G {\rm Tr}[ (\tens{c} - \tens{I}) \cdot \tens{V} ].
\end{equation}
The Rayleighian is
\begin{equation}
  \label{eq:R3}
  \mathscr{R} = G\tau {\rm Tr} [ \tens{c} \cdot ( \tens{V}^t - \tens{\kappa}^t) 
         \cdot ( \tens{V} - \tens{\kappa}) ]
     + G {\rm Tr}[ (\tens{c} - \tens{I}) \cdot \tens{V} ].
\end{equation}
Onsager principle requires $\partial \mathscr{R}/ \partial \tens{V} = 0$,
\begin{eqnarray}
  && 2G\tau ( \tens{V} - \tens{\kappa} ) \cdot \tens{c} + G ( \tens{c} - \tens{I} ) = 0, \\
  && \tens{V} = \tens{\kappa} - \frac{1}{2\tau} (\tens{I} - \tens{c}^{-1}).
  \label{eq:V2}
\end{eqnarray}

From Eqs.~(\ref{eq:cdot}) and (\ref{eq:V2}), we obtain the time evolution of $\tens{c}$
\begin{equation}
  \label{eq:cons3}
  \dot{\tens{c}} - \tens{\kappa} \cdot \tens{c}  - \tens{c} \cdot \tens{\kappa}^t = - \frac{1}{\tau} ( \tens{c} - \tens{I} ).
\end{equation}
Equations~(\ref{eq:R3}) and (\ref{eq:V2}) give the constitutive equation
\begin{equation}
  \tens{\sigma} = \frac{ \partial \mathscr{R} }{ \partial \tens{\kappa}} 
    = -G\tau (\tens{V} - \tens{\kappa}) \cdot \tens{c} = G (\tens{c}-\tens{I}).
\end{equation}

%=========================================
\subsection{Convenient form of the c-tensor formalism}

Let 
\begin{equation}
  \tens{V} = \tens{\kappa} + \tens{X} \cdot \tens{c}^{-1},
\end{equation}
From Eq.~(\ref{eq:cdotV}),
\begin{eqnarray}
  \dot{\tens{c}} = \tens{V} \cdot \tens{c} + \tens{c} \cdot \tens{V}^t 
       = \tens{\kappa} \cdot \tens{c} + \tens{X} + \tens{c} \cdot \tens{\kappa}^t + \tens{X}^t, \\
  \tens{X} + \tens{X}^t = \dot{\tens{c}} - \tens{\kappa} \cdot \tens{c} - \tens{c} \cdot \tens{\kappa}^t \equiv \hat{\tens{c}}.
\end{eqnarray}

It can be shown at the minimum, $\tens{X}$ is symmetric
\begin{equation}
  \tens{X} = \frac{1}{2} \hat{\tens{c}}, \quad 
  \tens{V} - \tens{\kappa} = \frac{1}{2} \hat{\tens{c}} \cdot \tens{c}^{-1}, \quad
  \tens{V}^t - \tens{\kappa}^t = \frac{1}{2} \tens{c}^{-1} \cdot \hat{\tens{c}}^t
\end{equation}

The dissipation (\ref{eq:Phi5}) is
\begin{equation}
  \Phi = \frac{G\tau}{4} {\rm Tr} ( \tens{c} \cdot \tens{c}^{-1} \cdot \hat{\tens{c}}^t 
         \cdot \hat{\tens{c}} \cdot \tens{c}^{-1} ) 
       = \frac{G\tau}{4} {\rm Tr} ( \hat{\tens{c}}^t \cdot \hat{\tens{c}} \cdot \tens{c}^{-1} )
\end{equation}
We finally arrive the final expression for the dissipation and the free energy
\begin{eqnarray}
  \label{eq:Phi0}
  \Phi &=& \frac{G\tau}{4} {\rm Tr} [ \tens{c}^{-1} ( \dot{\tens{c}}^t - \tens{\kappa} \cdot \tens{c} - 
         \tens{c} \cdot \tens{\kappa}^t) ( \dot{\tens{c}} - \tens{\kappa} \cdot \tens{c} - 
         \tens{c} \cdot \tens{\kappa}^t) ], \\
  A &=& \frac{1}{2} G [ {\rm Tr}(\tens{c}) - \ln \det(\tens{c}) ] \\
  \dot{A} &=& \frac{1}{2} G {\rm Tr} [ ( \tens{I} - \tens{c}^{-1}) \cdot \dot{\tens{c}} ] 
\end{eqnarray}

The Rayleighian is given by $\mathscr{R} = \Phi + \dot{A}$.
The evolution equations are given by $\partial \mathscr{R}/\partial \dot{\tens{c}} = 0$
\begin{equation}
  \dot{\tens{c}} - \tens{\kappa} \cdot \tens{c} - \tens{c} \cdot \tens{\kappa}^t 
    = - \frac{1}{\tau} (\tens{c}-\tens{I}).
\end{equation}
The stress tensor is given by 
\begin{equation}
  \tens{\sigma} = \frac{\partial \mathscr{R}}{\partial \tens{\kappa}}= G(\tens{c}-\tens{I}).
\end{equation}

%%%%%%%%%%%%%%%%%%%%%%%%%%%%%%%%%%%%%%%%%%%%%%%%%%%%%%%%%%%%%%%%
\section{Derivation of one-dimensional equations by Onsager principle}
\label{app:1dEqn}

In this section, we derive the one-dimensional evolution equations using Onsager principle. 
Our derivation is based on the Oldroyd-B model, assuming that the polymeric liquid column is described by the radius $h(z,t)$, and the polymer configuration $c_{rr}(z,t)$ and $c_{zz}(z,t)$, which are the radial and axial components, respectively.
The corresponding dynamic variables are $v(z,t)$, the averaged velocity in $z$-direction, and the material derivative of $\tens{c}$-tensor
\begin{equation}
  \label{eq:DcDt}
  \uD_t c_{\alpha\alpha} = \frac{\uD c_{\alpha\alpha}}{\uD t} = \frac{\partial c_{\alpha\alpha}}{\partial t} + v \frac{\partial c_{\alpha\alpha}}{\partial z}.
\end{equation}
%The material derivatives are used here because we now are working in a fixed-grid Lagrangian framework. 
In the discussion of Sec. \ref{sec:Uniform_stretching}, the $\tens{c}$-tensor is a constant along the $z$-axis, so $\uD_t c=\uD c/\uD t$ and $\partial c/\partial t$ coincide.

The original derivation of one-dimensional equations were given in Ref.~\cite{Forest1990} and quoted in Ref.~\cite{Clasen2006}. 
We list them here for reference, using our notations. 
\begin{eqnarray}
  \label{eq:1d_dh2dt}
  && \frac{\partial h^2}{\partial t} + \frac{\partial}{\partial z} (vh^2) = 0, \\
  \label{eq:1d_dvdt}
  && \frac{\partial v}{\partial t} + v \frac{\partial v}{\partial z} = \frac{1}{h^2} \frac{\partial}{\partial z} \left[ h^2 \left( K + 3 \eta_s \frac{\partial v}{\partial z} + \sigma^{(p)}_{zz} - \sigma^{(p)}_{rr} \right) \right], \\
  \label{eq:1d_K}
  && K = \gamma \Big( \frac{1}{h(1+h_z^2)^{1/2}} + \frac{h_{zz}}{(1+h_z^2)^{3/2}} \Big),  \\
  \label{eq:1d_dsigmazzdt}
  && \frac{\partial \sigma_{zz}^{(p)}}{\partial t} + v \frac{\partial \sigma_{zz}^{(p)}}{\partial z} = 2 \frac{\partial v}{\partial z} \sigma_{zz}^{(p)} + 2 G \frac{\partial v}{\partial z} - \sigma_{zz}^{(p)}, \\
  \label{eq:1d_dsigmaxxdt}
  && \frac{\partial \sigma_{rr}^{(p)}}{\partial t} + v \frac{\partial \sigma_{rr}^{(p)}}{\partial z} = - \frac{\partial v}{\partial z} \sigma_{rr}^{(p)} - G \frac{\partial v}{\partial z} - \sigma_{rr}^{(p)}.
\end{eqnarray}
The first equation (\ref{eq:1d_dh2dt}) is the continuity equation to ensure the volume conservation. 
The second equation (\ref{eq:1d_dvdt}) is the force balance equation, where $K$ [given by Eq.~(\ref{eq:1d_K})] is related to the surface curvature and $h_z$ indicates the partial derivative $\partial h/\partial z$.
The last two equations (\ref{eq:1d_dsigmazzdt}) and (\ref{eq:1d_dsigmaxxdt}) are the evolution equations for the polymer contribution to the stress tensor $\boldsymbol{\sigma}^{(p)}$; only the diagonal terms are considered here. 

We proceed with Onsager principle. 
The dissipation function is given by Eq.~(\ref{eq:el_Phi}) but the integration is performed in the cylindrical coordinates
\begin{equation}
  \Phi = \int \pi h^2  \left[ \frac{3}{2} \eta_s \Big( \frac{\partial v}{\partial z} \Big)^2  
           + \frac{\eta_p}{4c_{zz}} \Big( \uD_t {c}_{zz} - 2 \frac{\partial v}{\partial z} c_{zz} \Big)^2 
           + \frac{\eta_p}{2c_{rr}} \Big( \uD_t {c}_{rr} +   \frac{\partial v}{\partial z} c_{rr} \Big)^2  \right] \ud z.
  \label{eq:1d_Phi}
\end{equation}
The variation of the dissipation function is 
\begin{eqnarray}
  \delta \Phi &=& \int \pi \Bigg\{ \frac{\partial}{\partial z} \left[ h^2 \left( - 3 \eta_s \frac{\partial v}{\partial z}  
           + \eta_p \Big( \uD_t {c}_{zz} - 2 \frac{\partial v}{\partial z} c_{zz} \Big) 
           - \eta_p \Big( \uD_t {c}_{rr} +   \frac{\partial v}{\partial z} c_{rr} \Big) \right) \right] \Bigg\} \delta v \ud z \nonumber \\
  &+& \int \pi h^2 \left[ \frac{\eta_p}{2c_{zz}} \Big( \uD_t {c}_{zz} - 2 \frac{\partial v}{\partial z} c_{zz} \Big) \right] \delta (\uD_t {c}_{zz}) \ud z \nonumber \\
  &+&  \int \pi h^2 \left[ \frac{\eta_p}{ c_{rr}} \Big( \uD_t {c}_{rr} +   \frac{\partial v}{\partial z} c_{rr} \Big) \right] \delta (\uD_t {c}_{rr}) \ud z .
      \label{eq:delPhi}
\end{eqnarray}

The polymer contribution to the free energy is
\begin{equation}
  A_p = \int \pi h^2 \left[ \frac{1}{2} G \left( c_{zz} + 2c_{rr} - \ln c_{zz} c_{rr}^2 \right) \right].
\end{equation}
The time derivative of $A_p$ is
\begin{equation}
  \label{eq:1d_Acdot}
  \dot{A}_p = \int \pi \frac{\partial h^2}{\partial t} \left[ \frac{1}{2} G \left( c_{zz} + 2c_{rr} - \ln c_{zz} c_{rr}^2 \right) \right] \ud z 
      + \int \pi h^2 \frac{\partial}{\partial t} \left[ \frac{1}{2} G \left( c_{zz} + 2c_{rr} - \ln c_{zz} c_{rr}^2 \right) \right] \ud z .
\end{equation}
Substitute Eq.~(\ref{eq:1d_dh2dt}) into the first term on the RHS of Eq.~(\ref{eq:1d_Acdot}), and perform an integration by part
\begin{eqnarray}
  && \int \pi \frac{\partial h^2}{\partial t} \left[ \frac{1}{2} G \left( c_{zz} + 2c_{rr} - \ln c_{zz} c_{rr}^2 \right) \right] \ud z \nonumber \\
     &=& - \int \pi \frac{\partial (v h^2)}{\partial z} \left[ \frac{1}{2} G \left( c_{zz} + 2c_{rr} - \ln c_{zz} c_{rr}^2 \right) \right] \ud z \nonumber \\
     &=& \int \pi v h^2 \frac{1}{2} G \Big( 
          (1-\frac{1}{c_{zz}}) \frac{\partial c_{zz}}{\partial z}
       +  (2-\frac{2}{c_{rr}}) \frac{\partial c_{rr}}{\partial z} \Big) \ud z . \nonumber
\end{eqnarray}
Using Eq.~(\ref{eq:DcDt}), the second term on RHS of Eq.~(\ref{eq:1d_Acdot}) can be written as
\begin{eqnarray}
  && \int \pi h^2 \frac{\partial}{\partial t} \left[ \frac{1}{2} G \left( c_{zz} + 2c_{rr} - \ln c_{zz} c_{rr}^2 \right) \right] \ud z \nonumber \\
  &=&  \int \pi h^2  \left[ \frac{1}{2} G \Big( 
          (1-\frac{1}{c_{zz}}) \frac{\partial c_{zz}}{\partial t} 
       +  (2-\frac{2}{c_{rr}}) \frac{\partial c_{rr}}{\partial t} \Big) \right] \ud z \nonumber \\
  &=& \int \pi h^2  \left[ \frac{1}{2} G \Big( 
          (1-\frac{1}{c_{zz}}) (\uD_t {c}_{zz} - v \frac{\partial c_{zz}}{\partial z}) 
       +  (2-\frac{2}{c_{rr}}) (\uD_t {c}_{rr} - v \frac{\partial c_{rr}}{\partial z}) \Big) \right] \ud z . \nonumber
\end{eqnarray}
The final result is
\begin{equation}
  \dot{A}_p = \int \pi h^2  \left[ \frac{1}{2} G \Big( 
          (1-\frac{1}{c_{zz}}) \uD_t {c}_{zz} 
       +  (2-\frac{2}{c_{rr}}) \uD_t {c}_{rr} \Big) \right] \ud z.
\end{equation}
The variation of $\dot{A}_p$ is
\begin{equation}
  \label{eq:delAc}
  \delta \dot{A}_p = \int \pi h^2 \frac{1}{2} G (1-\frac{1}{c_{zz}}) \delta (\uD_t {c}_{zz}) \ud z 
                   + \int \pi h^2             G (1-\frac{1}{c_{rr}}) \delta (\uD_t {c}_{rr}) \ud z .
\end{equation}

The surface contribution to the free energy is
\begin{equation}
  A_{\gamma} = \gamma \int 2 \pi h (1+h_z^2)^{1/2} \ud z .
\end{equation}
The time derivative of $A_{\gamma}$ is
\begin{eqnarray}
  \dot{A}_{\gamma} &=& \gamma \int 2\pi \left[ \frac{\partial h}{\partial t} (1+h_z^2)^{1/2} + h h_z (1+h_z^2)^{-1/2} \frac{\partial}{\partial z} \Big( \frac{\partial h}{\partial t} \Big) \right] \ud z \nonumber \\
  &=& \gamma \int 2\pi \left[ \frac{\partial h}{\partial t} (1+h_z^2)^{1/2} + h h_z (1+h_z^2)^{-1/2} \frac{\partial}{\partial z} \Big( - \frac{1}{2h} \frac{\partial (v h^2)}{\partial z} \Big) \right] \ud z \nonumber \\
  &=& \gamma \int \left[ 2\pi \frac{\partial h}{\partial t} (1+h_z^2)^{1/2} + \pi h h_z (1+h_z^2)^{-1/2} \Big( \frac{h_z}{h^2} \frac{\partial (v h^2)}{\partial z} - \frac{1}{h} \frac{\partial^2 (v h^2)}{\partial z^2} \Big) \right] \ud z .
  \label{eq:Agam}
\end{eqnarray}

The first term in Eq. (\ref{eq:Agam}) is 
\begin{eqnarray}
  & &\gamma \int 2\pi \frac{\partial h}{\partial t} ( 1+h_z^2 )^{1/2} \ud z \nonumber \\
  &=& \gamma \int 2\pi \left( - \frac{1}{2h} \frac{\partial (vh^2)}{\partial z} \right) (1+h_z^2)^{1/2} \ud z \nonumber \\
  &=& - \gamma \int \pi \frac{ (1+h_z^2)^{1/2} }{h} \ud ( vh^2) \nonumber \\
  &=& \gamma \int \pi v h^2 \frac{\partial}{\partial z} \left( \frac{ (1+h_z^2)^{1/2} }{h} \right) \ud z \nonumber \\
  &=& \gamma \int \pi v \Big[ h h_z h_{zz} ( 1+h_z^2)^{-1/2} - h_z (1+h_z^2)^{1/2} \Big] \ud z
\end{eqnarray}
The second term in Eq. (\ref{eq:Agam}) is
\begin{eqnarray}
  & & \gamma \int \pi h h_z (1+h_z^2)^{-1/2} \Big( \frac{h_z}{h^2} \frac{\partial (v h^2)}{\partial z} \Big) \ud z \nonumber \\
  &=& \gamma \int \pi \left( \frac{ h_z^2 ( 1+h_z^2 )^{-1/2} }{h} \right) \ud (vh^2) \nonumber \\
  &=& - \gamma \int \pi v h^2 \frac{\partial}{\partial z} \left( \frac{ h_z^2 ( 1+h_z^2 )^{-1/2} }{h} \right) \ud z  \nonumber \\
  &=& \gamma \int \pi v \Big[ (h_z^3-2 h h_z h_{zz}) ( 1+h_z^2)^{-1/2} + h h_z^3 h_{zz} (1+h_z^2)^{-3/2} \Big] \ud z
\end{eqnarray}
The third term in Eq. (\ref{eq:Agam}) is
\begin{eqnarray}
  & & \gamma \int \pi h h_z (1+h_z^2)^{-1/2} \Big( - \frac{1}{h} \frac{\partial^2 (v h^2)}{\partial z^2} \Big) \ud z \nonumber \\
  &=& - \gamma \int \pi h_z (1+h_z^2)^{-1/2} \ud \left( \frac{\partial (vh^2)}{\partial z} \right) \nonumber \\
  &=&  \gamma \int \pi \frac{\partial (vh^2)}{\partial z} \frac{\partial}{\partial z} \left( h_z ( 1+h_z^2)^{-1/2} \right) \ud z \nonumber \\
  &=& \gamma \int \pi \Big( h_{zz}(1+h_z^2)^{-1/2} - h_z^2 h_{zz} (1+h_z^2)^{-3/2} \Big) \ud (vh^2) \nonumber \\
  &=& -\gamma \int \pi vh^2 \frac{\partial}{\partial z} \Big( h_{zz}(1+h_z^2)^{-1/2} - h_z^2 h_{zz} (1+h_z^2)^{-3/2} \Big) \ud z ] \nonumber \\
  &=& \gamma \int \pi v h^2 \Big[ -h_{zzz}(1+h_z^2)^{-1/2} + (3h_zh_{zz} + h_z^2h_{zzz})(1+h_z^2)^{-3/2} \nonumber \\
  && - 3h_z^3 h_{zz}^2 ( 1+h_z^2)^{-5/2} \Big] \ud z 
     \label{eq:dotAg}
\end{eqnarray}

Combining above three equations, we can obtain
\begin{eqnarray}
\dot{A}_{\gamma} & =& \gamma \int \pi v \Big[ - h_z (1+h_z^2)^{-1/2} + ( h_z^3 - h h_z h_{zz} - h^2 h_{zzz} ) (1+h_z^2)^{-3/2} \nonumber \\
  &&+ ( h h_z^3 h_{zz} + 3 h^2 h_z h_{zz}^2 + h^2 h_z^2 h_{zzz} ) (1+h_z^2)^{-3/2} - 3 h^2 h_z^3 h_{zz}^2 (1+h_z^2)^{-5/2} \Big] \ud z \nonumber \\
  &=& \gamma \int \pi v \Big[ -h_z - 2h_z^3 - h_z^5 - h h_z h_{zz} - h^2  h_{zzz} - h h_z^3 h_{zz} - h^2 h_z^2 h_{zzz} + 3h^2 h_z h_{zz}^2 \Big] (1+h_z^2)^{-5/2}  \ud z \nonumber \\
  &=& - \gamma \int \pi v  \Big[ h_z(1+h_z^2)^{-1/2} + (hh_zh_{zz} + h^2 h_{zzz}) (1+h_z^2)^{-3/2} - 3h^2 h_z h_{zz}^2 (1+h_z^2)^{-5/2} \Big] \ud z 
     \label{eq:K}
\end{eqnarray}

From Eq. (\ref{eq:1d_K}), we can get
\begin{eqnarray}
  \frac{\partial}{\partial z}(h^2K) &=& \gamma \frac{\partial}{\partial z} \Big[ h(1+h_z^2)^{-1/2} +  h h_{zz} (1+h_z^2)^{-3/2} \Big] \nonumber \\
  &=& \gamma \Big[ h_z(1+h_z^2)^{-1/2} + (hh_zh_{zz} + h^2 h_{zzz}) (1+h_z^2)^{-3/2} \nonumber \\
  && - 3h^2 h_z h_{zz}^2 (1+h_z^2)^{-5/2} \Big]. 
\end{eqnarray}
Comparing Eqs. (\ref{eq:dotAg}) and (\ref{eq:K}), we can arrive 
\begin{equation}
  \dot{A}_{\gamma} = - \int \pi v \frac{\partial (h^2 K)}{\partial z} \ud z .
\end{equation}

The variation of $\dot{A}_{\gamma}$ is
\begin{equation}
  \label{eq:delAg}
  \delta \dot{A}_{\gamma} = - \int \pi \frac{\partial (h^2 K)}{\partial z} \delta v \ud z.
\end{equation}

Using Eqs.~(\ref{eq:delAc}) and (\ref{eq:delAg}), we can write the variation of Rayleighian $\mathscr{R} = \Phi + \dot{A}_p + \dot{A}_{\gamma}$ with respect to $\uD_t {c}_{\alpha\alpha}$
\begin{eqnarray}
  \frac{\delta \mathscr{R}}{\delta (\uD_t {c}_{zz})} = 0 &\quad \Rightarrow \quad &
    \uD_t {c}_{zz} - 2 \frac{\partial v}{\partial z} c_{zz} + \frac{1}{\tau} (c_{zz}-1) = 0 , \\
  \frac{\delta \mathscr{R}}{\delta (\uD_t {c}_{rr})} = 0 &\quad \Rightarrow \quad &
    \uD_t {c}_{rr} +   \frac{\partial v}{\partial z} c_{rr} + \frac{1}{\tau} (c_{rr}-1) = 0 .
\end{eqnarray}
These can be rewritten in term of polymer contributions to the stress tensor $\sigma^{(p)}_{\alpha\alpha} = G(c_{\alpha\alpha}-1)$
\begin{eqnarray}
  \frac{\partial \sigma_{zz}^{(p)}}{\partial t} + v \frac{\partial \sigma_{zz}^{(p)}}{\partial z} = 2 \frac{\partial v}{\partial z} \sigma_{zz}^{(p)} + 2 G \frac{\partial v}{\partial z} - \sigma_{zz}^{(p)}, \\
  \frac{\partial \sigma_{rr}^{(p)}}{\partial t} + v \frac{\partial \sigma_{rr}^{(p)}}{\partial z} = - \frac{\partial v}{\partial z} \sigma_{rr}^{(p)} - G \frac{\partial v}{\partial z} - \sigma_{rr}^{(p)}.
\end{eqnarray}
These are the same as Eqs.~(\ref{eq:1d_dsigmazzdt}) and (\ref{eq:1d_dsigmaxxdt}).

Using Eq.~(\ref{eq:delPhi}), we can write the variation of Rayleighian $\mathscr{R} = \Phi + \dot{A}_p + \dot{A}_{\gamma}$ with respect to $v$
\begin{equation} 
   \frac{\delta \mathscr{R}}{\delta v} = 0 \quad \Rightarrow \quad 
     \frac{\partial}{\partial z} \left[h^2 \left( K + 3\eta_s \frac{\partial v}{\partial z} + \sigma_{zz}^{(p)} - \sigma_{rr}^{(p)} \right) \right] = 0.
\end{equation}
This equation agrees with Eq.~(\ref{eq:1d_dvdt}) if the inertial terms are negligible.

%==================================================
%==================================================
\bibliography{ve}

\begin{thebibliography}{37}%
\makeatletter
\providecommand \@ifxundefined [1]{%
 \@ifx{#1\undefined}
}%
\providecommand \@ifnum [1]{%
 \ifnum #1\expandafter \@firstoftwo
 \else \expandafter \@secondoftwo
 \fi
}%
\providecommand \@ifx [1]{%
 \ifx #1\expandafter \@firstoftwo
 \else \expandafter \@secondoftwo
 \fi
}%
\providecommand \natexlab [1]{#1}%
\providecommand \enquote  [1]{``#1''}%
\providecommand \bibnamefont  [1]{#1}%
\providecommand \bibfnamefont [1]{#1}%
\providecommand \citenamefont [1]{#1}%
\providecommand \href@noop [0]{\@secondoftwo}%
\providecommand \href [0]{\begingroup \@sanitize@url \@href}%
\providecommand \@href[1]{\@@startlink{#1}\@@href}%
\providecommand \@@href[1]{\endgroup#1\@@endlink}%
\providecommand \@sanitize@url [0]{\catcode `\\12\catcode `\$12\catcode
  `\&12\catcode `\#12\catcode `\^12\catcode `\_12\catcode `\%12\relax}%
\providecommand \@@startlink[1]{}%
\providecommand \@@endlink[0]{}%
\providecommand \url  [0]{\begingroup\@sanitize@url \@url }%
\providecommand \@url [1]{\endgroup\@href {#1}{\urlprefix }}%
\providecommand \urlprefix  [0]{URL }%
\providecommand \Eprint [0]{\href }%
\providecommand \doibase [0]{http://dx.doi.org/}%
\providecommand \selectlanguage [0]{\@gobble}%
\providecommand \bibinfo  [0]{\@secondoftwo}%
\providecommand \bibfield  [0]{\@secondoftwo}%
\providecommand \translation [1]{[#1]}%
\providecommand \BibitemOpen [0]{}%
\providecommand \bibitemStop [0]{}%
\providecommand \bibitemNoStop [0]{.\EOS\space}%
\providecommand \EOS [0]{\spacefactor3000\relax}%
\providecommand \BibitemShut  [1]{\csname bibitem#1\endcsname}%
\let\auto@bib@innerbib\@empty
%</preamble>
\bibitem [{\citenamefont {Bird}\ \emph
  {et~al.}(1987{\natexlab{a}})\citenamefont {Bird}, \citenamefont {Armstrong},\
  and\ \citenamefont {Hassager}}]{BAH1}%
  \BibitemOpen
  \bibfield  {author} {\bibinfo {author} {\bibfnamefont {R.~Byron}\
  \bibnamefont {Bird}}, \bibinfo {author} {\bibfnamefont {Robert~C.}\
  \bibnamefont {Armstrong}}, \ and\ \bibinfo {author} {\bibfnamefont {Ole}\
  \bibnamefont {Hassager}},\ }\href@noop {} {\emph {\bibinfo {title} {Dynamics
  of Polymeric Liquids Volume 1: Fluid Mechanics}}},\ \bibinfo {edition} {2nd}\
  ed.\ (\bibinfo  {publisher} {John Wiley \& Son},\ \bibinfo {year}
  {1987})\BibitemShut {NoStop}%
\bibitem [{\citenamefont {Bird}\ \emph
  {et~al.}(1987{\natexlab{b}})\citenamefont {Bird}, \citenamefont {Curtiss},
  \citenamefont {Armstrong},\ and\ \citenamefont {Hassager}}]{BAH2}%
  \BibitemOpen
  \bibfield  {author} {\bibinfo {author} {\bibfnamefont {R.~Byron}\
  \bibnamefont {Bird}}, \bibinfo {author} {\bibfnamefont {Charles~F.}\
  \bibnamefont {Curtiss}}, \bibinfo {author} {\bibfnamefont {Robert~C.}\
  \bibnamefont {Armstrong}}, \ and\ \bibinfo {author} {\bibfnamefont {Ole}\
  \bibnamefont {Hassager}},\ }\href@noop {} {\emph {\bibinfo {title} {Dynamics
  of Polymeric Liquids Volume 2: Kinetic Theory}}},\ \bibinfo {edition} {2nd}\
  ed.\ (\bibinfo  {publisher} {John Wiley \& Son},\ \bibinfo {year}
  {1987})\BibitemShut {NoStop}%
\bibitem [{\citenamefont {Yarin}\ \emph {et~al.}(2014)\citenamefont {Yarin},
  \citenamefont {Pourdeyhimi},\ and\ \citenamefont
  {Ramakrishna}}]{Yarin_fibers}%
  \BibitemOpen
  \bibfield  {author} {\bibinfo {author} {\bibfnamefont {Alexander~L.}\
  \bibnamefont {Yarin}}, \bibinfo {author} {\bibfnamefont {Behnam}\
  \bibnamefont {Pourdeyhimi}}, \ and\ \bibinfo {author} {\bibfnamefont
  {Seeram}\ \bibnamefont {Ramakrishna}},\ }\href@noop {} {\emph {\bibinfo
  {title} {Fundamentals and Applications of Micro- and Nanofibers}}}\ (\bibinfo
   {publisher} {Cambridge University Press},\ \bibinfo {address} {Cambridge,
  UK},\ \bibinfo {year} {2014})\BibitemShut {NoStop}%
\bibitem [{\citenamefont {Yarin}(1993)}]{Yarin}%
  \BibitemOpen
  \bibfield  {author} {\bibinfo {author} {\bibfnamefont {Alexander~L.}\
  \bibnamefont {Yarin}},\ }\href@noop {} {\emph {\bibinfo {title} {Free Liquid
  Jets and Films: Hydrodynamics and Rheology}}}\ (\bibinfo  {publisher}
  {Longman Scientific \& Technical},\ \bibinfo {year} {1993})\BibitemShut
  {NoStop}%
\bibitem [{\citenamefont {McKinley}\ and\ \citenamefont
  {Sridhar}(2002)}]{McKinley2002}%
  \BibitemOpen
  \bibfield  {author} {\bibinfo {author} {\bibfnamefont {Gareth~H.}\
  \bibnamefont {McKinley}}\ and\ \bibinfo {author} {\bibfnamefont {Tamarapu}\
  \bibnamefont {Sridhar}},\ }\bibfield  {title} {\enquote {\bibinfo {title}
  {Filament-stretching rheometry of complex fluids},}\ }\href {\doibase
  10.1146/annurev.fluid.34.083001.125207} {\bibfield  {journal} {\bibinfo
  {journal} {Annu. Rev. Fluid Mech.}\ }\textbf {\bibinfo {volume} {34}},\
  \bibinfo {pages} {375--415} (\bibinfo {year} {2002})}\BibitemShut {NoStop}%
\bibitem [{\citenamefont {McKinley}(2005)}]{McKinley2005}%
  \BibitemOpen
  \bibfield  {author} {\bibinfo {author} {\bibfnamefont {Gareth~H.}\
  \bibnamefont {McKinley}},\ }\bibfield  {title} {\enquote {\bibinfo {title}
  {Visco-elasto-capillary thinning and break-up of complex fluids},}\
  }\href@noop {} {\bibfield  {journal} {\bibinfo  {journal} {Rheol. Rev.}\
  }\textbf {\bibinfo {volume} {2005}},\ \bibinfo {pages} {1--48} (\bibinfo
  {year} {2005})}\BibitemShut {NoStop}%
\bibitem [{\citenamefont {Huang}\ and\ \citenamefont
  {Hassager}(2017)}]{HuangQian2017}%
  \BibitemOpen
  \bibfield  {author} {\bibinfo {author} {\bibfnamefont {Qian}\ \bibnamefont
  {Huang}}\ and\ \bibinfo {author} {\bibfnamefont {Ole}\ \bibnamefont
  {Hassager}},\ }\bibfield  {title} {\enquote {\bibinfo {title} {Polymer
  liquids fracture like solids},}\ }\href {\doibase 10.1039/c7sm00126f}
  {\bibfield  {journal} {\bibinfo  {journal} {Soft Matter}\ }\textbf {\bibinfo
  {volume} {13}},\ \bibinfo {pages} {3470--3474} (\bibinfo {year}
  {2017})}\BibitemShut {NoStop}%
\bibitem [{\citenamefont {Eggers}(1997)}]{Eggers1997}%
  \BibitemOpen
  \bibfield  {author} {\bibinfo {author} {\bibfnamefont {Jens}\ \bibnamefont
  {Eggers}},\ }\bibfield  {title} {\enquote {\bibinfo {title} {Nonlinear
  dynamics and breakup of free-surface flows},}\ }\href {\doibase
  10.1103/revmodphys.69.865} {\bibfield  {journal} {\bibinfo  {journal} {Rev.
  Mod. Phys.}\ }\textbf {\bibinfo {volume} {69}},\ \bibinfo {pages} {865--930}
  (\bibinfo {year} {1997})}\BibitemShut {NoStop}%
\bibitem [{\citenamefont {Goldin}\ \emph {et~al.}(1969)\citenamefont {Goldin},
  \citenamefont {Yerushalmi}, \citenamefont {Pfeffer},\ and\ \citenamefont
  {Shinnar}}]{Goldin1969}%
  \BibitemOpen
  \bibfield  {author} {\bibinfo {author} {\bibfnamefont {Michael}\ \bibnamefont
  {Goldin}}, \bibinfo {author} {\bibfnamefont {Joseph}\ \bibnamefont
  {Yerushalmi}}, \bibinfo {author} {\bibfnamefont {Robert}\ \bibnamefont
  {Pfeffer}}, \ and\ \bibinfo {author} {\bibfnamefont {Reuel}\ \bibnamefont
  {Shinnar}},\ }\bibfield  {title} {\enquote {\bibinfo {title} {Breakup of a
  laminar capillary jet of a viscoelastic fluid},}\ }\href {\doibase
  10.1017/s0022112069002540} {\bibfield  {journal} {\bibinfo  {journal} {J.
  Fluid Mech.}\ }\textbf {\bibinfo {volume} {38}},\ \bibinfo {pages} {689}
  (\bibinfo {year} {1969})}\BibitemShut {NoStop}%
\bibitem [{\citenamefont {Renardy}(1994)}]{Renardy1994}%
  \BibitemOpen
  \bibfield  {author} {\bibinfo {author} {\bibfnamefont {Michael}\ \bibnamefont
  {Renardy}},\ }\bibfield  {title} {\enquote {\bibinfo {title} {Some comments
  on the surface-tension driven break-up (or the lack of it) of viscoelastic
  jets},}\ }\href {\doibase 10.1016/0377-0257(94)85005-4} {\bibfield  {journal}
  {\bibinfo  {journal} {J. Non-Newtonian Fluid Mech.}\ }\textbf {\bibinfo
  {volume} {51}},\ \bibinfo {pages} {97--107} (\bibinfo {year}
  {1994})}\BibitemShut {NoStop}%
\bibitem [{\citenamefont {Renardy}(1995)}]{Renardy1995}%
  \BibitemOpen
  \bibfield  {author} {\bibinfo {author} {\bibfnamefont {Michael}\ \bibnamefont
  {Renardy}},\ }\bibfield  {title} {\enquote {\bibinfo {title} {A numerical
  study of the asymptotic evolution and breakup of newtonian and viscoelastic
  jets},}\ }\href {\doibase 10.1016/0377-0257(95)01375-6} {\bibfield  {journal}
  {\bibinfo  {journal} {J. Non-Newtonian Fluid Mech.}\ }\textbf {\bibinfo
  {volume} {59}},\ \bibinfo {pages} {267--282} (\bibinfo {year}
  {1995})}\BibitemShut {NoStop}%
\bibitem [{\citenamefont {Li}\ and\ \citenamefont
  {Fontelos}(2003)}]{LiJie2003}%
  \BibitemOpen
  \bibfield  {author} {\bibinfo {author} {\bibfnamefont {Jie}\ \bibnamefont
  {Li}}\ and\ \bibinfo {author} {\bibfnamefont {Marco~A.}\ \bibnamefont
  {Fontelos}},\ }\bibfield  {title} {\enquote {\bibinfo {title} {Drop dynamics
  on the beads-on-string structure for viscoelastic jets: A numerical study},}\
  }\href {\doibase 10.1063/1.1556291} {\bibfield  {journal} {\bibinfo
  {journal} {Phys. Fluids}\ }\textbf {\bibinfo {volume} {15}},\ \bibinfo
  {pages} {922} (\bibinfo {year} {2003})}\BibitemShut {NoStop}%
\bibitem [{\citenamefont {Wagner}\ \emph {et~al.}(2005)\citenamefont {Wagner},
  \citenamefont {Amarouchene}, \citenamefont {Bonn},\ and\ \citenamefont
  {Eggers}}]{Wagner2005}%
  \BibitemOpen
  \bibfield  {author} {\bibinfo {author} {\bibfnamefont {C.}~\bibnamefont
  {Wagner}}, \bibinfo {author} {\bibfnamefont {Y.}~\bibnamefont {Amarouchene}},
  \bibinfo {author} {\bibfnamefont {Daniel}\ \bibnamefont {Bonn}}, \ and\
  \bibinfo {author} {\bibfnamefont {J.}~\bibnamefont {Eggers}},\ }\bibfield
  {title} {\enquote {\bibinfo {title} {Droplet detachment and satellite bead
  formation in viscoelastic fluids},}\ }\href {\doibase
  10.1103/physrevlett.95.164504} {\bibfield  {journal} {\bibinfo  {journal}
  {Phys. Rev. Lett.}\ }\textbf {\bibinfo {volume} {95}},\ \bibinfo {pages}
  {164504} (\bibinfo {year} {2005})}\BibitemShut {NoStop}%
\bibitem [{\citenamefont {Clasen}\ \emph {et~al.}(2006)\citenamefont {Clasen},
  \citenamefont {Eggers}, \citenamefont {Fontelos}, \citenamefont {Li},\ and\
  \citenamefont {McKinley}}]{Clasen2006}%
  \BibitemOpen
  \bibfield  {author} {\bibinfo {author} {\bibfnamefont {Christian}\
  \bibnamefont {Clasen}}, \bibinfo {author} {\bibfnamefont {Jens}\ \bibnamefont
  {Eggers}}, \bibinfo {author} {\bibfnamefont {Marco~A.}\ \bibnamefont
  {Fontelos}}, \bibinfo {author} {\bibfnamefont {Jie}\ \bibnamefont {Li}}, \
  and\ \bibinfo {author} {\bibfnamefont {Gareth~H.}\ \bibnamefont {McKinley}},\
  }\bibfield  {title} {\enquote {\bibinfo {title} {The beads-on-string
  structure of viscoelastic threads},}\ }\href {\doibase
  10.1017/s0022112006009633} {\bibfield  {journal} {\bibinfo  {journal} {J.
  Fluid Mech.}\ }\textbf {\bibinfo {volume} {556}},\ \bibinfo {pages} {283}
  (\bibinfo {year} {2006})}\BibitemShut {NoStop}%
\bibitem [{\citenamefont {Sattler}\ \emph {et~al.}(2008)\citenamefont
  {Sattler}, \citenamefont {Wagner},\ and\ \citenamefont
  {Eggers}}]{Sattler2008}%
  \BibitemOpen
  \bibfield  {author} {\bibinfo {author} {\bibfnamefont {R.}~\bibnamefont
  {Sattler}}, \bibinfo {author} {\bibfnamefont {C.}~\bibnamefont {Wagner}}, \
  and\ \bibinfo {author} {\bibfnamefont {J.}~\bibnamefont {Eggers}},\
  }\bibfield  {title} {\enquote {\bibinfo {title} {Blistering pattern and
  formation of nanofibers in capillary thinning of polymer solutions},}\ }\href
  {\doibase 10.1103/physrevlett.100.164502} {\bibfield  {journal} {\bibinfo
  {journal} {Phys. Rev. Lett.}\ }\textbf {\bibinfo {volume} {100}},\ \bibinfo
  {pages} {164502} (\bibinfo {year} {2008})}\BibitemShut {NoStop}%
\bibitem [{\citenamefont {Entov}\ and\ \citenamefont
  {Hinch}(1997)}]{Entov1997}%
  \BibitemOpen
  \bibfield  {author} {\bibinfo {author} {\bibfnamefont {V.M.}\ \bibnamefont
  {Entov}}\ and\ \bibinfo {author} {\bibfnamefont {E.J.}\ \bibnamefont
  {Hinch}},\ }\bibfield  {title} {\enquote {\bibinfo {title} {Effect of a
  spectrum of relaxation times on the capillary thinning of a filament of
  elastic liquid},}\ }\href {\doibase 10.1016/s0377-0257(97)00022-0} {\bibfield
   {journal} {\bibinfo  {journal} {J. Non-Newtonian Fluid Mech.}\ }\textbf
  {\bibinfo {volume} {72}},\ \bibinfo {pages} {31--53} (\bibinfo {year}
  {1997})}\BibitemShut {NoStop}%
\bibitem [{\citenamefont {Beris}\ and\ \citenamefont
  {Edwards}(1994)}]{Beris_Edwards}%
  \BibitemOpen
  \bibfield  {author} {\bibinfo {author} {\bibfnamefont {Antony~N.}\
  \bibnamefont {Beris}}\ and\ \bibinfo {author} {\bibfnamefont {Brian~J.}\
  \bibnamefont {Edwards}},\ }\href@noop {} {\emph {\bibinfo {title}
  {Thermodynamics of Flowing Systems}}}\ (\bibinfo  {publisher} {Oxford
  University Press},\ \bibinfo {address} {Oxford},\ \bibinfo {year}
  {1994})\BibitemShut {NoStop}%
\bibitem [{\citenamefont {{\"O}ttinger}(2005)}]{Oettinger_beyond}%
  \BibitemOpen
  \bibfield  {author} {\bibinfo {author} {\bibfnamefont {Hans~Christian}\
  \bibnamefont {{\"O}ttinger}},\ }\href@noop {} {\emph {\bibinfo {title}
  {Beyond Equilibrium Thermodynamics}}}\ (\bibinfo  {publisher} {John Wiley \&
  Son},\ \bibinfo {address} {New Jersey},\ \bibinfo {year} {2005})\BibitemShut
  {NoStop}%
\bibitem [{\citenamefont {Doi}(2013)}]{DoiSoft}%
  \BibitemOpen
  \bibfield  {author} {\bibinfo {author} {\bibfnamefont {Masao}\ \bibnamefont
  {Doi}},\ }\href@noop {} {\emph {\bibinfo {title} {Soft Matter Physics}}}\
  (\bibinfo  {publisher} {Oxford University Press},\ \bibinfo {year}
  {2013})\BibitemShut {NoStop}%
\bibitem [{\citenamefont {Onsager}(1931)}]{Onsager1931}%
  \BibitemOpen
  \bibfield  {author} {\bibinfo {author} {\bibfnamefont {Lars}\ \bibnamefont
  {Onsager}},\ }\bibfield  {title} {\enquote {\bibinfo {title} {Reciprocal
  relations in irreversible processes. i.}}\ }\href {\doibase
  10.1103/physrev.37.405} {\bibfield  {journal} {\bibinfo  {journal} {Phys.
  Rev.}\ }\textbf {\bibinfo {volume} {37}},\ \bibinfo {pages} {405--426}
  (\bibinfo {year} {1931})}\BibitemShut {NoStop}%
\bibitem [{\citenamefont {Doi}(2012)}]{Doi2012}%
  \BibitemOpen
  \bibfield  {author} {\bibinfo {author} {\bibfnamefont {Masao}\ \bibnamefont
  {Doi}},\ }\bibfield  {title} {\enquote {\bibinfo {title} {Onsager's
  variational principle in soft matter dynamics},}\ }in\ \href@noop {} {\emph
  {\bibinfo {booktitle} {Non-Equilibrium Soft Matter Physics}}},\ \bibinfo
  {editor} {edited by\ \bibinfo {editor} {\bibfnamefont {S.}~\bibnamefont
  {Komura}}\ and\ \bibinfo {editor} {\bibfnamefont {T.}~\bibnamefont {Ohta}}}\
  (\bibinfo  {publisher} {World Scientific},\ \bibinfo {year} {2012})\ pp.\
  \bibinfo {pages} {1--35}\BibitemShut {NoStop}%
\bibitem [{\citenamefont {Forest}\ and\ \citenamefont
  {Wang}(1990)}]{Forest1990}%
  \BibitemOpen
  \bibfield  {author} {\bibinfo {author} {\bibfnamefont {M.~Gregory}\
  \bibnamefont {Forest}}\ and\ \bibinfo {author} {\bibfnamefont
  {Qi}~\bibnamefont {Wang}},\ }\bibfield  {title} {\enquote {\bibinfo {title}
  {Change-of-type behavior in viscoelastic slender jet models},}\ }\href
  {\doibase 10.1007/bf00271426} {\bibfield  {journal} {\bibinfo  {journal}
  {Theor. Comput. Fluid Dyn.}\ }\textbf {\bibinfo {volume} {2}},\ \bibinfo
  {pages} {1--25} (\bibinfo {year} {1990})}\BibitemShut {NoStop}%
\bibitem [{\citenamefont {Doi}(2015)}]{Doi2015}%
  \BibitemOpen
  \bibfield  {author} {\bibinfo {author} {\bibfnamefont {Masao}\ \bibnamefont
  {Doi}},\ }\bibfield  {title} {\enquote {\bibinfo {title} {Onsager principle
  as a tool for approximation},}\ }\href {\doibase
  10.1088/1674-1056/24/2/020505} {\bibfield  {journal} {\bibinfo  {journal}
  {Chin. Phys. B}\ }\textbf {\bibinfo {volume} {24}},\ \bibinfo {pages}
  {020505} (\bibinfo {year} {2015})}\BibitemShut {NoStop}%
\bibitem [{\citenamefont {Doi}(2016)}]{Doi2016}%
  \BibitemOpen
  \bibfield  {author} {\bibinfo {author} {\bibfnamefont {Masao}\ \bibnamefont
  {Doi}},\ }\bibfield  {title} {\enquote {\bibinfo {title} {A principle in
  dynamic coarse graining{\textendash}{Onsager} principle and its
  applications},}\ }\href {\doibase 10.1140/epjst/e2016-60128-5} {\bibfield
  {journal} {\bibinfo  {journal} {Eur. Phys. J. Special Topics}\ }\textbf
  {\bibinfo {volume} {225}},\ \bibinfo {pages} {1411} (\bibinfo {year}
  {2016})}\BibitemShut {NoStop}%
\bibitem [{\citenamefont {Tomari}\ and\ \citenamefont
  {Doi}(1995)}]{Tomari1995}%
  \BibitemOpen
  \bibfield  {author} {\bibinfo {author} {\bibfnamefont {Tsutomu}\ \bibnamefont
  {Tomari}}\ and\ \bibinfo {author} {\bibfnamefont {Masao}\ \bibnamefont
  {Doi}},\ }\bibfield  {title} {\enquote {\bibinfo {title} {Hysteresis and
  incubation in the dynamics of volume transition of spherical gels},}\ }\href
  {\doibase 10.1021/ma00128a050} {\bibfield  {journal} {\bibinfo  {journal}
  {Macromolecules}\ }\textbf {\bibinfo {volume} {28}},\ \bibinfo {pages}
  {8334--8343} (\bibinfo {year} {1995})}\BibitemShut {NoStop}%
\bibitem [{\citenamefont {Meng}\ \emph {et~al.}(2016)\citenamefont {Meng},
  \citenamefont {Luo}, \citenamefont {Doi},\ and\ \citenamefont
  {Ouyang}}]{MengFanlong2016a}%
  \BibitemOpen
  \bibfield  {author} {\bibinfo {author} {\bibfnamefont {Fanlong}\ \bibnamefont
  {Meng}}, \bibinfo {author} {\bibfnamefont {Ling}\ \bibnamefont {Luo}},
  \bibinfo {author} {\bibfnamefont {Masao}\ \bibnamefont {Doi}}, \ and\
  \bibinfo {author} {\bibfnamefont {Zhongcan}\ \bibnamefont {Ouyang}},\
  }\bibfield  {title} {\enquote {\bibinfo {title} {Solute based lagrangian
  scheme in modeling the drying process of soft matter solutions},}\ }\href
  {\doibase 10.1140/epje/i2016-16022-9} {\bibfield  {journal} {\bibinfo
  {journal} {Eur. Phys. J. E}\ }\textbf {\bibinfo {volume} {39}},\ \bibinfo
  {pages} {22} (\bibinfo {year} {2016})}\BibitemShut {NoStop}%
\bibitem [{\citenamefont {Man}\ and\ \citenamefont
  {Doi}(2016)}]{ManXingkun2016}%
  \BibitemOpen
  \bibfield  {author} {\bibinfo {author} {\bibfnamefont {Xingkun}\ \bibnamefont
  {Man}}\ and\ \bibinfo {author} {\bibfnamefont {Masao}\ \bibnamefont {Doi}},\
  }\bibfield  {title} {\enquote {\bibinfo {title} {Ring to mountain transition
  in deposition pattern of drying droplets},}\ }\href {\doibase
  10.1103/PhysRevLett.116.066101} {\bibfield  {journal} {\bibinfo  {journal}
  {Phys. Rev. Lett.}\ }\textbf {\bibinfo {volume} {116}},\ \bibinfo {pages}
  {066101} (\bibinfo {year} {2016})}\BibitemShut {NoStop}%
\bibitem [{\citenamefont {Di}\ \emph {et~al.}(2016)\citenamefont {Di},
  \citenamefont {Xu},\ and\ \citenamefont {Doi}}]{DiYana2016}%
  \BibitemOpen
  \bibfield  {author} {\bibinfo {author} {\bibfnamefont {Yana}\ \bibnamefont
  {Di}}, \bibinfo {author} {\bibfnamefont {Xianmin}\ \bibnamefont {Xu}}, \ and\
  \bibinfo {author} {\bibfnamefont {Masao}\ \bibnamefont {Doi}},\ }\bibfield
  {title} {\enquote {\bibinfo {title} {Theoretical analysis for meniscus rise
  of a liquid contained between a flexible film and a solid wall},}\ }\href
  {\doibase 10.1209/0295-5075/113/36001} {\bibfield  {journal} {\bibinfo
  {journal} {Europhys. Lett.}\ }\textbf {\bibinfo {volume} {113}},\ \bibinfo
  {pages} {36001} (\bibinfo {year} {2016})}\BibitemShut {NoStop}%
\bibitem [{\citenamefont {Xu}\ \emph {et~al.}(2016)\citenamefont {Xu},
  \citenamefont {Di},\ and\ \citenamefont {Doi}}]{XuXianmin2016}%
  \BibitemOpen
  \bibfield  {author} {\bibinfo {author} {\bibfnamefont {Xianmin}\ \bibnamefont
  {Xu}}, \bibinfo {author} {\bibfnamefont {Yana}\ \bibnamefont {Di}}, \ and\
  \bibinfo {author} {\bibfnamefont {Masao}\ \bibnamefont {Doi}},\ }\bibfield
  {title} {\enquote {\bibinfo {title} {Variational method for contact line
  problems in sliding liquids},}\ }\href {\doibase 10.1063/1.4959227}
  {\bibfield  {journal} {\bibinfo  {journal} {Phys. Fluids}\ }\textbf {\bibinfo
  {volume} {28}},\ \bibinfo {pages} {087101} (\bibinfo {year}
  {2016})}\BibitemShut {NoStop}%
\bibitem [{\citenamefont {Zhou}\ \emph {et~al.}(2017)\citenamefont {Zhou},
  \citenamefont {Jiang},\ and\ \citenamefont {Doi}}]{2017_stratification}%
  \BibitemOpen
  \bibfield  {author} {\bibinfo {author} {\bibfnamefont {Jiajia}\ \bibnamefont
  {Zhou}}, \bibinfo {author} {\bibfnamefont {Ying}\ \bibnamefont {Jiang}}, \
  and\ \bibinfo {author} {\bibfnamefont {Masao}\ \bibnamefont {Doi}},\
  }\bibfield  {title} {\enquote {\bibinfo {title} {Cross interaction drives
  stratification in drying film of binary colloidal mixtures},}\ }\href
  {\doibase 10.1103/PhysRevLett.118.108002} {\bibfield  {journal} {\bibinfo
  {journal} {Phys. Rev. Lett.}\ }\textbf {\bibinfo {volume} {118}},\ \bibinfo
  {pages} {108002} (\bibinfo {year} {2017})}\BibitemShut {NoStop}%
\bibitem [{\citenamefont {Mora}\ \emph {et~al.}(2010)\citenamefont {Mora},
  \citenamefont {Phou}, \citenamefont {Fromental}, \citenamefont {Pismen},\
  and\ \citenamefont {Pomeau}}]{Mora2010}%
  \BibitemOpen
  \bibfield  {author} {\bibinfo {author} {\bibfnamefont {Serge}\ \bibnamefont
  {Mora}}, \bibinfo {author} {\bibfnamefont {Ty}~\bibnamefont {Phou}}, \bibinfo
  {author} {\bibfnamefont {Jean-Marc}\ \bibnamefont {Fromental}}, \bibinfo
  {author} {\bibfnamefont {Len~M.}\ \bibnamefont {Pismen}}, \ and\ \bibinfo
  {author} {\bibfnamefont {Yves}\ \bibnamefont {Pomeau}},\ }\bibfield  {title}
  {\enquote {\bibinfo {title} {Capillarity driven instability of a soft
  solid},}\ }\href {\doibase 10.1103/physrevlett.105.214301} {\bibfield
  {journal} {\bibinfo  {journal} {Phys. Rev. Lett.}\ }\textbf {\bibinfo
  {volume} {105}},\ \bibinfo {pages} {214301} (\bibinfo {year}
  {2010})}\BibitemShut {NoStop}%
\bibitem [{\citenamefont {Xuan}\ and\ \citenamefont
  {Biggins}(2016)}]{XuanChen2016}%
  \BibitemOpen
  \bibfield  {author} {\bibinfo {author} {\bibfnamefont {Chen}\ \bibnamefont
  {Xuan}}\ and\ \bibinfo {author} {\bibfnamefont {John}\ \bibnamefont
  {Biggins}},\ }\bibfield  {title} {\enquote {\bibinfo {title}
  {Finite-wavelength surface-tension-driven instabilities in soft solids,
  including instability in a cylindrical channel through an elastic solid},}\
  }\href {\doibase 10.1103/PhysRevE.94.023107} {\bibfield  {journal} {\bibinfo
  {journal} {Phys. Rev. E}\ }\textbf {\bibinfo {volume} {94}},\ \bibinfo
  {pages} {023107} (\bibinfo {year} {2016})}\BibitemShut {NoStop}%
\bibitem [{\citenamefont {Anna}\ and\ \citenamefont
  {McKinley}(2001)}]{Anna2001}%
  \BibitemOpen
  \bibfield  {author} {\bibinfo {author} {\bibfnamefont {Shelley~L.}\
  \bibnamefont {Anna}}\ and\ \bibinfo {author} {\bibfnamefont {Gareth~H.}\
  \bibnamefont {McKinley}},\ }\bibfield  {title} {\enquote {\bibinfo {title}
  {Elasto-capillary thinning and breakup of model elastic liquids},}\ }\href
  {\doibase 10.1122/1.1332389} {\bibfield  {journal} {\bibinfo  {journal} {J.
  Rheol.}\ }\textbf {\bibinfo {volume} {45}},\ \bibinfo {pages} {115} (\bibinfo
  {year} {2001})}\BibitemShut {NoStop}%
\bibitem [{\citenamefont {Xuan}\ and\ \citenamefont
  {Biggins}(2017)}]{XuanChen2017}%
  \BibitemOpen
  \bibfield  {author} {\bibinfo {author} {\bibfnamefont {Chen}\ \bibnamefont
  {Xuan}}\ and\ \bibinfo {author} {\bibfnamefont {John}\ \bibnamefont
  {Biggins}},\ }\bibfield  {title} {\enquote {\bibinfo {title}
  {{Plateau-Rayleigh} instability in solids is a simple phase separation},}\
  }\href {\doibase 10.1103/PhysRevE.95.053106} {\bibfield  {journal} {\bibinfo
  {journal} {Phys. Rev. E}\ }\textbf {\bibinfo {volume} {95}},\ \bibinfo
  {pages} {053106} (\bibinfo {year} {2017})}\BibitemShut {NoStop}%
\bibitem [{\citenamefont {Stelter}\ \emph {et~al.}(2000)\citenamefont
  {Stelter}, \citenamefont {Brenn}, \citenamefont {Yarin}, \citenamefont
  {Singh},\ and\ \citenamefont {Durst}}]{Stelter2000}%
  \BibitemOpen
  \bibfield  {author} {\bibinfo {author} {\bibfnamefont {M.}~\bibnamefont
  {Stelter}}, \bibinfo {author} {\bibfnamefont {G.}~\bibnamefont {Brenn}},
  \bibinfo {author} {\bibfnamefont {A.~L.}\ \bibnamefont {Yarin}}, \bibinfo
  {author} {\bibfnamefont {R.~P.}\ \bibnamefont {Singh}}, \ and\ \bibinfo
  {author} {\bibfnamefont {F.}~\bibnamefont {Durst}},\ }\bibfield  {title}
  {\enquote {\bibinfo {title} {Validation and application of a novel
  elongational device for polymer solutions},}\ }\href {\doibase
  10.1122/1.551102} {\bibfield  {journal} {\bibinfo  {journal} {J. Rheol.}\
  }\textbf {\bibinfo {volume} {44}},\ \bibinfo {pages} {595} (\bibinfo {year}
  {2000})}\BibitemShut {NoStop}%
\bibitem [{\citenamefont {Bazilevskii}\ and\ \citenamefont
  {Rozhkov}(2014)}]{Bazilevskii2014}%
  \BibitemOpen
  \bibfield  {author} {\bibinfo {author} {\bibfnamefont {A.~V.}\ \bibnamefont
  {Bazilevskii}}\ and\ \bibinfo {author} {\bibfnamefont {A.~N.}\ \bibnamefont
  {Rozhkov}},\ }\bibfield  {title} {\enquote {\bibinfo {title} {Dynamics of
  capillary breakup of elastic jets},}\ }\href {\doibase
  10.1134/s0015462814060143} {\bibfield  {journal} {\bibinfo  {journal} {Fluid
  Dynamics}\ }\textbf {\bibinfo {volume} {49}},\ \bibinfo {pages} {827--843}
  (\bibinfo {year} {2014})}\BibitemShut {NoStop}%
\bibitem [{\citenamefont {Bazilevskii}\ and\ \citenamefont
  {Rozhkov}(2015)}]{Bazilevskii2015}%
  \BibitemOpen
  \bibfield  {author} {\bibinfo {author} {\bibfnamefont {A.~B.}\ \bibnamefont
  {Bazilevskii}}\ and\ \bibinfo {author} {\bibfnamefont {A.~N.}\ \bibnamefont
  {Rozhkov}},\ }\bibfield  {title} {\enquote {\bibinfo {title} {Dynamics of the
  capillary breakup of a bridge in an elastic fluid},}\ }\href {\doibase
  10.1134/s0015462815060101} {\bibfield  {journal} {\bibinfo  {journal} {Fluid
  Dynamics}\ }\textbf {\bibinfo {volume} {50}},\ \bibinfo {pages} {800--811}
  (\bibinfo {year} {2015})}\BibitemShut {NoStop}%
\end{thebibliography}%
%\bibliography{/Users/zhouj2/bib/polymer}
%\bibliography{/home/zhou/bib/polymer}

% uncomment the following lines if using preprint endfloats
%\listoffigures

\end{document}